\documentclass{aa}  

\usepackage{graphicx}
\usepackage{txfonts}
\usepackage{lipsum}
\usepackage{subcaption}
\usepackage{lscape}
\usepackage{placeins}
\usepackage{amsmath}
\usepackage{amssymb}
\usepackage{upgreek}
\usepackage{xspace}
\usepackage{derivative}
\usepackage{xcolor}
\usepackage{hyperref}

\newcommand{\lcdm}{$\Lambda$CDM\xspace}
\newcommand{\OpenGadget}{\texttt{OpenGadget3}\xspace}
\newcommand{\pGadgetThree}{\texttt{P-Gadget3}\xspace}
\newcommand{\pGadgetTwo}{\texttt{GADGET-2}\xspace}
\newcommand{\MGMAMPOSSt}{\texttt{MG-MAMPOSSt}\xspace}

\begin{document}

   \title{Searching for signatures of self-interacting dark matter in halos from full physics simulations: From 3D structure to projected observables}

   \author{G. Y. Ferron\inst{1}\thanks{\email{g.ferron2@campus.unimib.it}}
        \and A. Ragagnin\inst{2}
        \and L. Pizzuti\inst{1,2}
        \and M.~S.~Fischer\inst{3}
        }

   \institute{Università di Milano Bicocca, Piazza della Scienza 3, 20126 Milan, Italy
   \and
   INAF-Osservatorio Astronomico di Trieste, Via G. B. Tiepolo 11, 34143 Trieste, Italy
   \and
   Donostia International Physics Center (DIPC), Paseo Manuel de Lardizabal 4, 20018 Donostia-San Sebastian, Spain
   }

   \date{Received 3 July 2026 / Accepted 24 August 2026}
 
  \abstract
 {
    In this work we explore the imprints of self-interacting dark matter (SIDM) on the structure of simulated galaxy groups and clusters in view of future optical surveys.
    We find that baryonic and projection effects both lead to a significant reduction of the SIDM core-forming behavior in the mass and density profiles, resulting in very small ($\lesssim 5\%$) deviations from the profiles of standard collisionless dark matter.
    The opposite trend emerged for our most massive halos, where SIDM significantly enhances the central density instead. We compared these halos with the observed dark matter profiles of two galaxy clusters, MACS~J1206 and Abell~S1063, finding that SIDM may leave potentially detectable imprints on cluster cores when considering the small observational uncertainty of current kinematic mass measurements.
    Finally, we find that weak lensing observations provide a promising tool for testing SIDM on cluster scales. In an idealized scenario where systematics are under control, weak lensing shear profiles exhibit deviations of up to $\sim 20\%$ for $M_{200} \gtrsim 10^{13}\ \text{M}_\odot$ halos.
    Conversely, the projected concentration--mass relation is only slightly modified by our SIDM models.

    We discuss the limitations of our analysis and the impact of the simulation resolution as well as possible extension of this work.
    Overall, our results provide a baseline for future SIDM studies. Although line-of-sight projection and baryonic effects significantly reduce SIDM imprints on halo mass profiles, weak lensing and kinematic analyses in galaxy clusters may potentially allow for robust detection and constraining of dark matter self-interactions in massive structures from megaparsec down to kiloparsec scales.  
    }


   \titlerunning{Influence of SIDM and baryons on 3D and 2D halo profiles}
   \authorrunning{G.~Y.~Ferron et al.}
   \maketitle
   \nolinenumbers

\section{Introduction}
Since its discovery in the early 1930s (\citealt{Zwicky_1933}, and~\citealt{Zwicky_2009} for a translation), the nature of dark matter (DM) has been a major problem in cosmology. The standard $\Lambda$-cold-dark-matter (\lcdm) paradigm assumes DM to be collisionless (CDM), interacting only through gravity and dominating over the energy density of ordinary matter, though its true nature and origin remain an open issue.

Many viable candidate particles for DM come from theories beyond the Standard Model of particle physics (see~\citealt{Bozorgnia_2025} for a review). The candidate particles only interact gravitationally on large scales, mimicking CDM, but at small scales they often allow for additional interactions strong enough that they should have a detectable impact on the internal structure of DM halos.
In this context, self-interacting dark matter (SIDM) was originally introduced by~\citet{Spergel_2000} as an alternative to CDM at small scales in order to solve the "core-cusp" and "missing satellites" problems (e.g.,~\citealt{de_Blok_2009, Oman_2015, Bullock_2017}). 
For a more in depth analysis on the vast array of SIDM models and their observational signatures one can refer to~\citet{Tulin_2018, Adhikari_2025}.

More generally, SIDM has been found to change various halo properties from dwarf galaxy (e.g.,~\citealt{Vogelsberger_2014}) up to galaxy cluster (e.g.,~\citealt{Robertson_2019, Sabarish_2024}) scales. 
These effects are only prevalent in the inner halo since as the density decreases outward, collisions become rare and SIDM more closely resembles CDM. In this way SIDM models recover the observationally favored predictions of CDM at large scales~\citep[e.g.,][]{Planck_2020} while also preserving the standard \lcdm cosmic growth history (e.g.,~\citealt{Spergel_2000}).

The past few years have seen an increasing number of studies focused on possible SIDM probes spanning many different observables~\citep{Adhikari_2025}, for example, the field of Milky Way satellites~\citep{Correa_2021,Yang_2023b,Shah_2024,Roberts_2025,Correa_2025,Nadler_2025,Silverman_2025,Roberts_2025,Ando_2025}; ultra-faint ~\citep{Almeida_2024, Fischer_Yu_2026} or diffuse galaxies~\citep{Zhang_2025b, Wang_2026} and strong galaxy-galaxy lensing perturbers~\citep{Despali_2022,Vegetti_2024,Lei_2025,Zhang_2025,Yu_2026}; strong lensing in galaxy cluster cores~\citep{Meneghetti_2020,Ragagnin_2022,Meneghetti_2022,Meneghetti_2023,Dutra_2025,ODonnell_2026,Natarajan_2026}; core buoyancy~\citep{vanDenBosch2026}; DM deficient galaxies~\cite{Wang_2026}; 3D central density of halos~\citep{Straight_2025}; brightest cluster galaxy dynamics~\citep{Harvey_2025}; possible subhalo matter redistribution and stripping~\citep{Sirks_2022,Tokayer_2024,Chiang_2026}; galaxy cluster mergers~\citep{Kim_2017, Robertson_2017, Fischer_2023, Sabarish_2024}; and baryon and DM misalignment~\citep{Sirks_2024}.

In this paper we partake in the investigation of potential SIDM probes by performing a novel search for signatures of DM self-interactions on weak lensing and kinematical properties of halo profiles inside groups and clusters of galaxies.
We are motivated in particular by the SIDM impact on weak lensing projected profiles since future cluster weak lensing mass reconstructions are expected to be sufficiently accurate to allow detection of very small SIDM deviations (e.g.,\ from \textit{Euclid} data stacked profiles~\citealt{EuclidSkyOverview,EP-Racz,Ingoglia_2025}). 
Moreover, mass reconstruction techniques based on the kinematics of member galaxies and their combination with gravitational lensing ~\citep{Mamon_2013,Pizzuti_2021,Umetsu25} can provide high-fidelity internal DM slope reconstructions~\citep{Biviano_2023,Pedratti_2026}.

Within this context, this work focuses on analyzing profile-based properties of the DM distribution inside simulated groups and clusters of galaxies in the presence of DM self-interactions in order to evaluate which halo profile observables are best suited for effective detection of SIDM imprints. 
At the time of writing, many theoretical studies tackling SIDM models have focused exclusively on the case of DM-only (DMO) and semi-analytic simulations or three-dimensional halo properties (e.g.,~\citealt{Ray_2022, Ray_2025, Romanello_2026, Engelhardt_2026}). For this reason, we utilized the SIDM hydrodynamical simulations of galaxy clusters presented in~\cite{Ragagnin_2024}, aiming to characterize how the interplay between the baryonic (gas plus stellar) component and SIDM affects halo profiles and how projected quantities are to be interpreted in view of upcoming observational campaigns such as \textit{Euclid} and DESI~\citep{DESI_Collaboration_2022}. In particular, we performed a stacking analysis of the 3D and 2D DM mass and density halo profiles, comparing our results between DMO and full physics simulations and providing an estimate of the halo weak lensing shear in the presence of SIDM by computing surface density excess profiles. For the most massive halos, we further discuss how SIDM imprints in our theoretical predictions compare to real galaxy profile observational uncertainties
through direct comparison with the results of~\cite{Biviano_2023} and~\cite{Sartoris_2020} , where multicomponent mass profiles of the massive clusters MACS~J1206 and Abell~S1063 from the CLASH-VLT survey~\citep{Postman_2012,Rosati2014} have been generated. We then briefly investigate how the concentration-mass (c-M) relation is affected by SIDM models when projected along the line of sight.

The paper is organized as follows: In Sect.~\ref{sec:SIDM_models} we provide an overview of the specific classes of SIDM implemented in the simulations, which are further discussed in Sect.~\ref{sec:DIANOGA_sim}. Section~\ref{sec:results} is devoted to the presentation of the principal results, while Sect.~\ref{sec:conc} provides a summary of our main conclusions.
Additional information is provided in the appendix.

\section{SIDM models} \label{sec:SIDM_models}

A wide range of SIDM models have been proposed in the literature (see~\citealt{Tulin_2018, Adhikari_2025} for reviews). In this work, we focus on two representative classes defined by the angular structure of the differential scattering cross section per unit mass. If scattering is approximately isotropic in the center-of-mass frame, collisions lead to an efficient exchange of momentum between particles. In this regime, even a relatively low interaction rate is sufficient to alter the internal structure of DM halos. We refer to this class as rare SIDM (rSIDM). Conversely, if the scattering is highly anisotropic and dominated by small deflection angles, the momentum transfer per interaction is suppressed. As a result, a significantly higher collision rate is required to produce comparable effects on halo profiles. This regime is commonly referred to as frequent SIDM (fSIDM; see, e.g.,~\citealt{Kahlhoefer_2014}).

In addition, the models considered here assume a velocity-dependent cross section. This choice is motivated by the difficulty of reconciling constraints across different astrophysical scales with a velocity-independent interaction. In particular, observations indicate a velocity dependence in order to abide by cross section constraints from galaxy clusters~\citep{Kaplinghat_2016, Silverman_2023, Lovell_2023}.

In this paper we employ the framework implemented in~\citep{Fischer_2021}, which is motivated by models where self-interactions arise from the exchange of a light scalar mediator.
In this case, the DM interaction can be described by a radial Yukawa potential~\citep{Loeb_2011, Tulin_2018}. Then, assuming elastic scattering, such a potential leads to a differential cross section, $\mathrm{d}\sigma/\mathrm{d}\Omega$, that generally depends on both the relative velocity of the colliding DM particles, $v_\mathrm{rel}$, and on the scattering angle in the center of mass frame, $\theta$~\citep{Girmohanta_2022}.
In the presence of anisotropic scattering, a useful way of expressing the significance of self-interactions is through the modified momentum-transfer cross section~\citep{Arido_2025}, defined as

\begin{equation} \label{eq:momentum_cross}
    \sigma_{\rm T} = 4 \uppi \int_0^\uppi{\odv{\sigma}{\Omega} (1 - |\! \cos{\theta}|) \sin{\theta} \ {\rm d} \theta}\, ,
\end{equation}

\noindent which is an angle-averaged cross section weighted by the forward momentum exchanged in collisions of identical particles. 
In general, rSIDM models are then associated with a differential cross section with typically large scattering angles. Meanwhile, fSIDM models represent the limit of very small scattering angles while also keeping the modified momentum transfer cross section, Eq.~\eqref{eq:momentum_cross}, constant. 

\begin{figure}[ht!]
    \centering
    \includegraphics[width=\hsize]{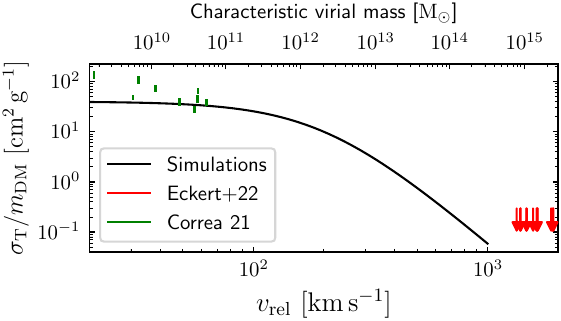}
    \caption{Momentum transfer cross section for the SIDM models considered in our study, given by Eq.~\eqref{eq:momentum_cross} as a black solid line. We overplot constrains from \cite{Eckert_2022} as red arrows and from \cite{Correa_2021} as green lines.
    The top axis reports the characteristic virial mass for halos following an NFW profile and a c-M relation from~\cite{Ludlow_2014}.}
    \label{fig:SIDM_cross}
\end{figure}

In terms of velocity dependence, for the specific case implemented in~\citet{Fischer_2024a} the transfer cross section assumes the following functional form:

\begin{equation} \label{eq:t_cross}
    \frac{\sigma_{\rm T} (v_{\rm rel})}{m_{\rm DM}} = \frac{\sigma_{\rm T, 0}}{m_{\rm DM}} \left[ 1 + \left( \frac{v_{\rm rel}}{v_{\rm c}} \right)^2 \right]^{-2}\, ,
\end{equation}
where the normalization $\sigma_{\rm T, 0}$ corresponds to the value of a low-velocity plateau of the cross section and $v_{\rm c}$ represents the transition velocity after which the cross section falls as $v_{\rm rel}^{-4}$. In this work we utilize a suite of SIDM simulations (Sect.~\ref{sec:DIANOGA_sim}) that set the normalization to $\sigma_{\rm T, 0} / m_{\rm DM} = 40\ \text{cm}^2 \, \text{g}^{-1}$ and $v_{\rm c} = 200\ \text{km} \, \text{s}^{-1}$, which are taken to be similar to the parameters presented in ~\citealt[][]{Correa_2021} (see their fig.~6).
Thus, with the functional form of Eq.~\eqref{eq:t_cross}, the cross section decreases with increasing DM particle velocity, consequently affecting low-mass halos most prominently while leaving large-scale cluster-sized objects mostly unaltered (see Fig.~\ref{fig:SIDM_cross}).

When choosing the cross sections for our simulations, we then matched the isotropic rSIDM and highly anisotropic fSIDM models using the modified transfer cross section due to historical reasons. Therefore, as we chose the same momentum transfer cross sections for both rSIDM and fSIDM, this led to fSIDM having a stronger effect on isolated halos, i.e.,\ they evolve faster. On the other hand, it is the viscosity cross section\footnote{The viscosity cross section is given by
$\sigma_v = 4\uppi \int_0^1 \frac{\mathrm{d}\sigma}{\mathrm{d}\Omega} \sin^2 \theta \, \mathrm{d}\cos\theta\, .$}
that better defines the evolution of a halo, and for this reason some of the differences between the models can be attributed to the match done using the transfer cross section. Indeed, in our simulations the fSIDM cross section is 50\% stronger compared to a match based on the viscosity cross section~\citep[see][]{Yang_2022, Sabarish_2024}. We note however that while the viscosity cross section works well to match the evolution of an isolated halo, the evolution of a halo in the cosmological context is more complicated. For example, during mergers it is less clear which cross section matches the evolution best, and \citep{Arido_2025} even showed -- for an idealized merger of two halos -- that the quality of the match depends on the quantity considered for the matching.

\section{The SIDM simulations} \label{sec:DIANOGA_sim}

To study the effects of the SIDM models presented in Sect.~\ref{sec:SIDM_models} we utilized the same suite of Dianoga 1x SIDM simulations first presented in~\citealt{Ragagnin_2024}.
This suite includes a number of zoom-in regions~\citep[see, e.g.,][]{Katz_1994, Navarro_1994} of the same parent box of comoving length 1~Gpc from initial conditions as used in~\cite{Bonafede_2011}\footnote{These initial conditions have been widely used to study galaxy and galaxy cluster properties~\citep{Rasia_2015,RagoneFigueroa_2018,Meneghetti_2020}.}
, who assumed a \lcdm initial power spectrum with cosmological parameters of $\Omega_{\rm m} = 0.24$, $\Omega_{\rm b} = 0.0456$, $n_{\rm s} = 0.96$, $\sigma_8 = 0.8$, and $h = 0.72$~\citep{Bonafede_2011}.
The DM mass resolution was set to $m_{DM} \simeq 1.3 \times 10^9\ \text{M}_\odot$ for all simulations, and the comoving softening parameter was $\epsilon_{\rm DM} = 3.7 h^{-1}\ \text{kpc}$ for DM and $\epsilon_{*} = 2.0h^{-1}\ \text{kpc}$ for the stellar component. 

The simulations were evolved using the code \OpenGadget\citep[see, e.g.,][for details]{Groth_2023,Sala_2024,Damiano_2024,TedeschiPrades_2025,Chaitra_2026,Dolag_2026}\footnote{\url{https://www.space-coe.eu/codes/opengadget.php}}, which is derived from the \pGadgetThree algorithm, an upgraded version of \pGadgetTwo~\citep{Springel_2005}.
The simulations followed DM and gas, with the latter simulated using a smoothed particle hydrodynamics~\citep[SPH; see][]{Price_2012,Dehnen_Aly_2012} improved solver~\citep{Dolag_2005,Beck_2016}. The solver also includes baryonic subgrid physics, such as star formation~\citep{Springel_Hernquist_2003}, cooling~\citep{Wiersma_2009,Tornatore_2010},\footnote{The cooling tables from the photo-ionization code CLOUDY can be found in~\cite{Ferland_1998}.} stellar evolution~\citep{Tornatore_2004,Tornatore_2007}, and thermal conduction~\citep{Dolag_2004}.
Moreover, \OpenGadget implements supermassive black hole growth and active galactic nucleus feedback based one the Magneticum simulations~\citep{Dolag_2025}. For more details on the black hole seeding and active galactic nucleus modeling, we refer the reader to~\citet{Springel_DiMatteo_2005,Fabjan_2010,Hirschmann_2014}. The DM self-interactions were implemented following the prescriptions in
\cite{Fischer_2021,Fisher_2022b,Fischer_2022,Fischer_2024a,Fischer_2026}, which, for example, guarantees consistent conservation of the kinetic energy during the scattering step.

For this study we considered six simulated zoom-in regions, each associated with a different galaxy cluster encompassing masses of $M_{200} / \text{M}_\odot \in \left[ 10^{14},\ 10^{15} \right]$\footnote{Note that $M_\Delta$ refers to the halo mass contained within a spherical shell of radius denoted as $r_{\Delta c}$, which encloses a mean halo density equal to the critical density of the Universe times $\Delta$. In particular, in this work we use $\Delta = 200$ and $\Delta=500,$ which correspond to masses of $M_{200}$ and $M_{500}$ and radii of $r_{200c}$ and $r_{500c}$, respectively.} (see Table~\ref{tab:DIANOGA_regions}).
Every region was re-simulated with a CDM, rSIDM, and fSIDM model, both as a DMO run and including baryons (denoted as "FP" for "full physics"). All host halos and their members were identified using the friends-of-friends (FoF) halo finder~\citep{Davis_1985} along with an upgraded version of the SubFind subhalo finder designed to account for the presence of baryons~\citep{Dolag_2009}. Furthermore, for the purposes of this work, we considered a single simulation snapshot at redshift $z = 0.4$, which is close to the redshift of the massive galaxy clusters MACS~J1206~\citep{Biviano_2023} and Abell~S1063~\citep{Mercurio_2021}, which we use as a benchmark for comparison with our simulated profiles in Sect.~\ref{sec:obs_cluster}.

\begin{table}[ht!]
\caption{Dark matter $M_{200}$ of the most massive halo in each DIANOGA-SIDM FP region analyzed in this work.
}              
\label{tab:DIANOGA_regions}    
\centering                        
\begin{tabular}{c c}      
\hline\hline              
Region & $M_{200}$ [$10^{14}\ \text{M}_\odot$]\\         
\hline                      
   D3 & 4.7\\    
   D4 & 2.3\\
   D5 & 1.2\\
   D10 & 8.3\\
   D15 & 7.5\\
   D16 & 9.7\\
\hline                                  
\end{tabular}
\tablefoot{We only show the cluster masses in CDM since their values are very similar ($\lesssim 2\%$ deviation) to the SIDM cases.}
\end{table}

The mass and density profiles were generated by dividing the particles into 50 logarithmically spaced radial bins inside the range $r/r_{500c} \in \left[ 10^{-3},\ 5 \right]$ for both 3D and projected profiles. Specifically, the projected profiles were computed along each Cartesian axis ($x, y, z$) to provide three independent lines of sight; we only considered particles within a cylinder of half-length $5 r_{500c}$ as a compromise between maximizing the radial coverage of our profiles while minimizing contamination from unrelated large-scale structure. 

\section{Results}
\label{sec:results}
In this section, we compare the effects of SIDM on DM profiles with and without the presence of baryons. In particular, we present the stacked mass, density, and surface density excess profiles across different mass scales, and we compare our results to real data from the MACS~J1206 and Abell~S1063 galaxy clusters. Finally, we conclude by briefly studying the median c-M relation.

\subsection{Halo mass fits} \label{sec:mass_fits}
Following our computation of the mass and density halo profiles, we performed fits of their cumulative mass profiles to obtain estimates of the halo scale radius, concentration, and virial mass. We assumed the widely used Navarro--Frenk--White (NFW) halo density model, which has been shown to provide a good match to the radial matter distribution inside simulated CDM halos over several scales ~\citep{Navarro_1996, Navarro_1997}.
The 3D NFW density profile can be defined as

\begin{equation} \label{eq:NFW_profile}
    \rho_{\rm NFW}(r) = \frac{\rho_0}{x (1 + x)^2} \quad , \quad x \equiv \frac{r}{r_{\rm s}}\, ,
\end{equation}

where $r_{\rm s}$ is the halo scale radius and $\rho_0$ is a characteristic density of the halo related to $r_{\rm s}$ and to the radius enclosing a mean halo density of 200 times the critical density of the Universe, $r_{200c}$. Thus, in this work we  used the NFW profile as a function of $r_{\rm s}$ and $r_{200c}$, which we consider as free parameters.

Being a differential quantity, the density profile is particularly affected by numerical noise given by the finite number of particles in each radial bin. Therefore, to perform a more robust fit we instead considered the corresponding cumulative mass profile, which we obtained by integrating the densities (Eq.~\ref{eq:NFW_profile}):

\begin{equation}
M(< r) = \int_0^r{4\uppi \rho(r')r'^2\ \text{d}r'}\,.
\end{equation}

We first fit the 3D NFW mass profiles using the method of nonlinear least squares, considering the Poissonian noise associated with each bin as uncertainties for the data points. We only considered particles within a selected radial interval, $r/r_{500c} \in \left[ 0.075,\ 1.5 \right]$, for both 3D and 2D cases. After performing several preliminary tests, we concluded that this range ensures reliable capture of the effects of SIDM in the inner halo while also avoiding complications associated with the inner core and outskirts. These 3D fits acted as a quality check of the particle-based halo finder $M_{200}$ estimates, which we later used for binning the stacked profiles, and they also provided an estimate of the concentration parameter of our simulated halos for the study of the halo c-M relation (Sect.~\ref{sec:c_m_relation}).

We further considered a projected NFW profile -- obtained by integrating the density profile (Eq.~\ref{eq:NFW_profile}) after projection along the line of sight -- to fit to the two-dimensional mass distribution. This 2D fit provided estimates of the c-M relation that more closely resemble real weak lensing measurements involving quantities projected onto the sky. 

In Appendix~\ref{app:mass_fits} we show ten randomly selected example halos in the \lcdm simulation also showing the distribution of best-fit $M_{200}$ compared to the \OpenGadget halo finder (FoF/SubFind) estimates. Overall, the distributions seem to be quite similar for all cosmologies.

Finally, by computing the halo virial mass and concentration we gained further insight into the SIDM core-formation process as a function of halo mass, through the estimation of the gravothermal evolution timescale for our simulated SIDM halos. For this purpose, we computed the halo central densities and compared them to the theoretical model of~\citep{Yang_2023}, as explained in detail in Appendix~\ref{app:grav_evolution}.

\subsection{Stacked mass and density profiles} \label{sec:stacked_profiles}

\begin{figure}[ht!]
    \centering
    \includegraphics[width=\hsize]{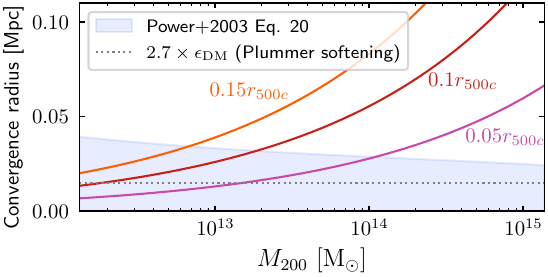}
    \caption{Convergence radius in megaparsec as a function of halo $M_{200}$ (shaded gray region).
    The solid curves indicate, for a given halo $M_{200}$, the value in megaparsec corresponding to a selected fraction of $r_{500c}$ (in this case 0.05, 0.1, and 0.15$r_{500c}$). We also display the resolution limit associated purely with the simulation DM gravitational softening, based on smoothing the gravitational force on scales smaller than the Plummer sphere scale radius~\citep{Plummer_1911}, i.e.,\ $\sim 2.7\epsilon_\mathrm{DM}$.}
    \label{fig:Power_resolution}
\end{figure}

To identify the typical SIDM imprints on the halo internal matter distribution, we performed a stacking analysis of the halo profiles, obtaining a statistically robust estimator of the effects of SIDM and baryons as a function of radius by extracting mean and median profile-based quantities. However, simulation resolution limits entailed that not all halos would be included in the analysis. Since lower-mass halos contain a smaller number of particles ($\lesssim 10^{3}$), they are particularly affected by numerical noise, and thus we could not guarantee reliable profile predictions. Therefore, to provide a robust analysis of SIDM profiles, we performed a mass cut at $M_{200} = 5 \times 10^{12}\ {\rm M_\odot}$ and considered only higher-mass halos. Furthermore, the cores of such higher-mass halos are also prone to being affected by resolution effects -- as they typically only contain $\lesssim 100$ particles -- resulting in a weak sampling of the inner profiles. In particular,~\cite{Power_2003} have shown that the mass contained within the inner halo regions heavily depends on the simulation resolution, thus requiring a high number of particles to be accurately resolved. They identified the minimum radius above which a given halo mass profile can be robustly traced with the radius at which the average DM collisional relaxation time approximately equals 0.6 times the age of the Universe (see Eq.~20 in~\citealt{Power_2003}). Therefore, our stacked profile analysis only reliably holds above that radius, while any eventual SIDM behavior present in the region below it might be significantly altered by resolution effects -- especially for differential quantities such as the density -- leading to artificial flattening of our profiles. Figure~\ref{fig:Power_resolution} shows the convergence radius as a function of halo mass, implying that we can only properly resolve the mass profiles of our low-mass halos down to $\sim 0.1 r_{500c}$. Finally we note that while SIDM profiles typically converge more easily compared to CDM ones, as the strength of the self-interactions is more important than the two-body gravitational interactions that artificially lower the central halo density (e.g.,~\citealt{Vogelsberger_2012}), we kept the same convergence radius for all of our DM models as a conservative estimate of the scale below which resolution is significant.

We also investigated the average profiles across different mass scales, dividing the halo sample into two different virial mass intervals. The interval edge was chosen to be the median of the CDM $M_{200}$ distribution in order to ensure roughly equally populated mass bins, resulting in the following mass intervals containing $\sim 200$ halos each: $M_{200} / \text{M}_\odot \in \left[ 5 \times 10^{12},\ 10^{13} \right)$ and $M_{200} / \text{M}_\odot \in \left[10^{13},\ 10^{15} \right]$. The bins are the same for the DMO and FP simulations since the contribution of baryons to the halo mass is negligible.

\begin{figure*}[ht!]
    \centering
    \centerline{
    \includegraphics[width=0.5\hsize]{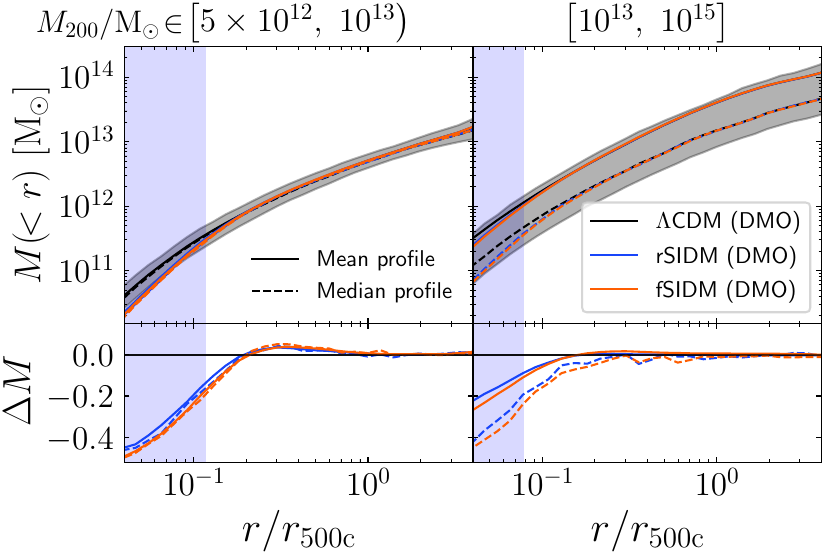}
    \includegraphics[width=0.5\hsize]{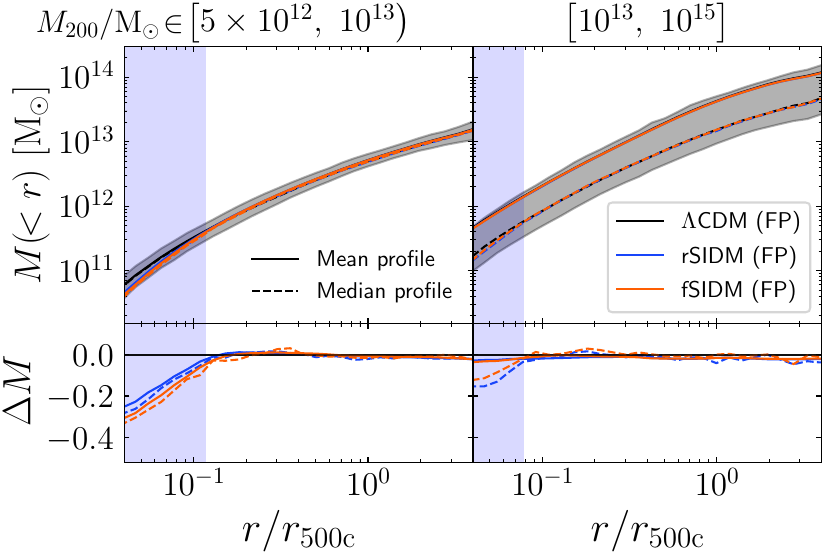}}
    \centerline{
    \includegraphics[width=0.5\hsize]{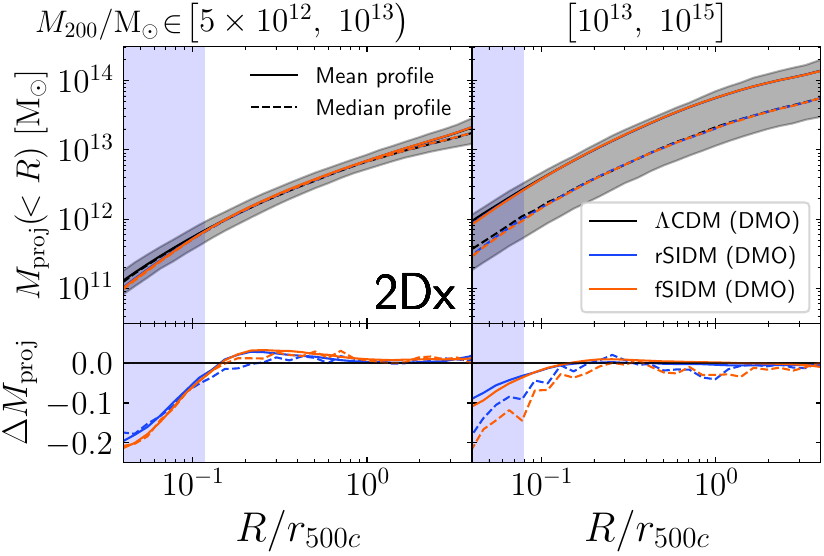}
    \includegraphics[width=0.5\hsize]{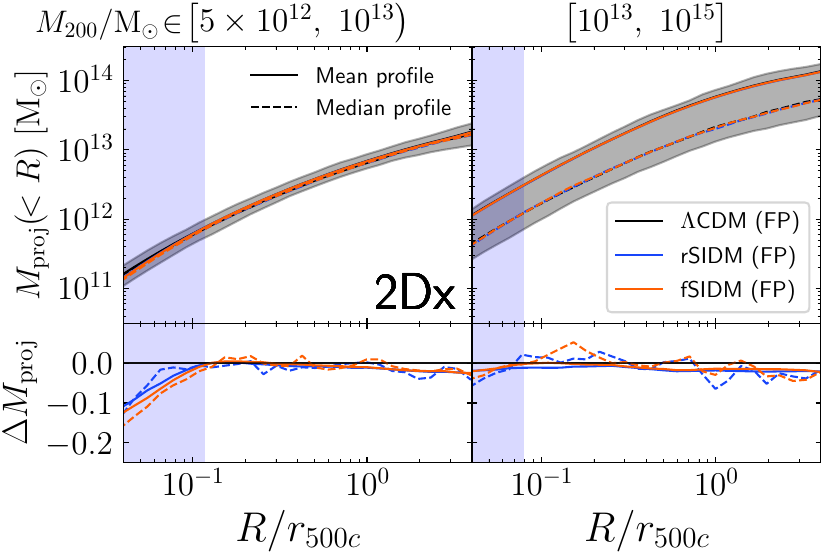}}
    \caption{Stacked cumulative DM mass profiles for the DMO (\textit{left}) and FP (\textit{right}) simulations in 3D (\textit{top}) and 2D (\textit{bottom}). The solid and dashed lines represent the mean and median profiles, respectively, and the line color is associated with the same DM model. We also show the $68\%$ scatter of the CDM model (gray shaded region) and the radial interval below which the simulation resolution plays a prominent role (blue shaded region), described in Sect.~\ref{sec:stacked_profiles}. The bottom subpanels show the relative differences between each model for both the mean and median profiles, where we define for a given quantity $X$, $\Delta X \equiv (X - X_{\rm CDM}) / X_{\rm CDM}$. Overall, SIDM halos feature a characteristic drop in the central profile and a mass bump just outside of it. Notably, while the shape of the projected profiles stays similar to the 3D case, they feature a less prominent mass drop in the center, most likely as a consequence of projection effects.}
    \label{fig:mass_profiles}
\end{figure*}

Figure~\ref{fig:mass_profiles} shows the 3D stacked mass profiles for all the DMO (left) and FP (right) simulations. The corresponding density profiles are reported in Appendix~\ref{app:density_profiles}. We computed both the mean (solid lines) and median (dashed lines) profiles as well as the relative differences between each SIDM model with respect to \lcdm, as shown in the bottom plots. The blue shaded regions indicate the radial intervals where the simulation mass resolution may significantly affect our profile estimates (as discussed at the beginning of this subsection) and were computed using the median $M_{200}$ of each mass bin. Overall, both SIDM models are characterized by a drop in the halo inner density -- reaching deviations from CDM of up to $10\% - 20\%$ above the convergence radius in the 3D mass DMO profiles -- which is a product of the SIDM collisions facilitating heat transfer across the halo, leading to more rarified inner regions than with CDM.
The addition of baryons resulted in a significant increase in the halo central mass and density as a result of the baryonic gas cooling and clustering in the center of the DM potential well, events associated with the presence of a brightest cluster galaxy enhancing the central concentration. This increase in baryon density affects the DM profile because the central gravitational potential becomes strong enough to pull DM particles into the core in a process called adiabatic contraction, creating a much steeper DM inner density profile (see, e.g.,~\citealt{Blumenthal_1986, Duffy_2010}). Indeed, although SIDM collisions act to oppose this process when the inner velocity dispersion gradient is positive and thus result in a shallower inner density profile than CDM on average, the DM distribution is still significantly affected, generally leading to much cuspier profiles than the DMO case. 

Although the outer halo profile is overall only slightly affected by self-interactions as the DM collision rate drops when density decreases, the mass profiles present a small ($\sim 5\%$) enhancement at $r/r_{500c} \sim 0.3 - 0.4$ in the DMO simulations at lower masses, which is absent in the high-mass bin.
This mass bump is a much more prominent feature of the density profiles, being found at the $\sim 5\% - 10\%$ level for all mass scales in both DMO and FP cases -- in agreement with previous and recent studies on SIDM profiles (e.g.,~\citealt{Ragagnin_2024, Despali_2026}). 
The origin of this small mass enhancement is related to the central SIDM heat conduction, which increases the inner velocity dispersion, diffusing core DM particles toward larger radii. The presence of baryon adiabatic contraction then opposes this heat transfer, suppressing the mass excess. Moreover, due to the declining relation of the SIDM cross section with velocity, the bump is also suppressed in the higher-mass bin.
Finally, we note the mismatch between mean and median profiles in the most massive bin caused by the presence of the few ($\lesssim 30$) very massive cluster-sized halos with $M_{200} \gtrsim 10^{14} - 10^{15}\ \text{M}_\odot$, which bias the mean estimate while leaving the median profile unaltered. In this case, the lack of very massive halos in our sample precludes us from conducting a proper stacking analysis at $M_{200} > 10^{14}\ {\rm M_\odot}$, so we included such halos alongside less massive ones in the highest mass bin and studied the SIDM behavior of the individual profiles for only three of our most massive halos in Sect.~\ref{sec:obs_cluster}.

\begin{figure}[ht!]
    \centering
    \includegraphics[width=0.95\hsize]{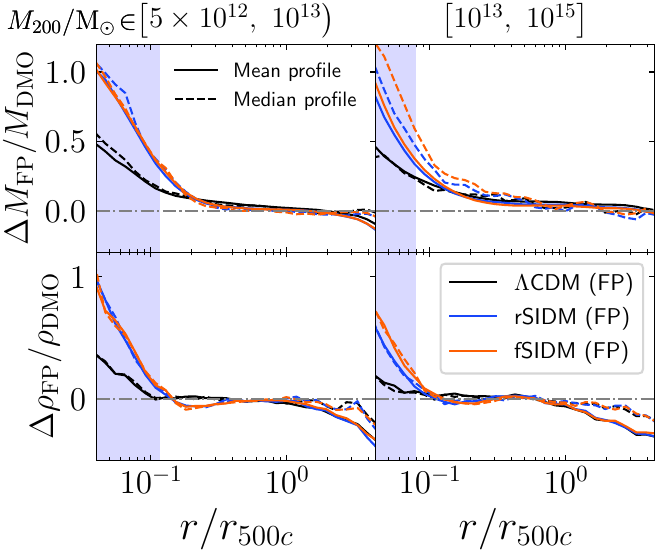}
    \caption{Relative deviations of the stacked FP profiles of DM mass and density compared to the DMO cases, where for a given quantity $X$, we define $\Delta X_{\rm FP} \equiv X_{\rm FP} - X_{\rm DMO}$. The color code is the same as in Fig.~\ref{fig:mass_profiles}.}
    \label{fig:3D_profiles_div}
\end{figure}

Figure~\ref{fig:3D_profiles_div} shows the relative differences between the stacked FP mass and density profiles and their DMO counterpart. The presence of a baryonic component at these mass scales enhances the central mass and density profiles more significantly when considering SIDM -- with fSIDM exhibiting greater enhancement at higher halo masses -- thus bringing the internal SIDM profiles much closer to CDM predictions (as shown in, e.g.,~\citealt{Robertson_2019, Despali_2026}). Interestingly, such behavior leads to cuspier FP SIDM profiles than FP CDM when considering the few highest mass halos ($M_{200} \sim 10^{15}\ \mathrm{M_\odot}$), as discussed in Sect.~\ref{sec:obs_cluster}. Conversely, at large radii baryons produce a suppression of the profiles consistently across all DM models, as SIDM behaves similar to CDM due to the low DM particle density. 

We also report in Fig.~\ref{fig:mass_profiles} (bottom panel) and in Appendix~\ref{app:density_profiles} the projected mass and density profiles along the $x$-direction (similar results hold for the other projections), which are a closer match to real-world scenarios, correlating directly to the mass and density profiles inferred from weak lensing observations (see Sect.~\ref{sec:lensing}). We find that projecting the profiles results in weaker SIDM behavior compared to their 3D counterparts, exhibiting a less pronounced decrease in the DMO central density ($\sim 5\% - 10\%$ above the resolution limit). Coupled with the presence of baryons, SIDM only produces at most $\lesssim 10\%$ deviations at the projected radius $R \sim 0.1 r_{500c}$, becoming indistinguishable from CDM at intermediate radii. However, in this case we note a further baryon-induced SIDM mass profile suppression at $R \gtrsim 0.5 - 0.6 r_{500c}$, at the $1\% - 2\%$ level. Overall, SIDM does not seem to generate any particular deviations above the resolution threshold in the 2D FP case; however, 2D profiles are known to be affected by various projection effects, which may alter our estimates significantly (Sect. \ref{sec:lensing}).

Finally, we also briefly mention that computation of the stacked mass and density profiles in lower halo mass bins ($M_{200} < 5 \times 10^{12}\ \mathrm{M_\odot}$) results in profiles very similar to those found for the higher-mass halos we presented. Nevertheless, a major difference is that low-mass halos are more affected by baryons in the CDM case rather than in the SIDM case (in contrast to the behavior of high-mass halos), resulting in boosted SIDM deviations from CDM in the FP case when compared to DMO. While this may confirm that the behavior of halo profiles described so far remains consistent at all mass scales, such low-mass halos are greatly affected by the simulation resolution, and therefore they cannot be used for reliable investigation on the actual magnitude of SIDM deviations. Thus, we do not consider them again for the purposes of this work. 

\subsection{Surface density excess profiles} \label{sec:lensing}
As discussed in the previous sections, 2D projection effects heavily wash out the SIDM signatures present in the three-dimensional halo profiles. Since weak gravitational lensing probes the projected matter distribution, it provides a natural observational framework to assess whether these imprints remain detectable once realistic line-of-sight effects are included.

In this context, we further investigated the weak lensing signal generated by our simulated halos -- assuming an idealized scenario where all the observational systematics are neglected -- by computing their surface density excess profiles. This quantity directly determines the tangential shear induced on background galaxies and therefore represents one of the main observables employed in current and upcoming lensing surveys. In particular, stacked weak lensing analyses of thousands of objects from ongoing surveys such as \textit{Euclid}~\citep{EuclidSkyOverview} may provide sufficiently precise projected mass reconstructions to test small deviations in halo profile shapes induced by SIDM.

The surface density excess profile is defined as

\begin{equation}
    \Delta \Sigma \equiv \Sigma(< R) - \Sigma(R)\, ,
\end{equation}

\noindent where $\Sigma(<R)$ is the mean surface density $\Sigma(R)$ contained within a ring of projected radius $R$.
Indeed, the weak lensing distortion generated by a spherically symmetric halo at redshift $z_l$ on background sources at redshift $z_s$ can be related to its projected density profile through the reduced tangential shear, a lensing observable given by

\begin{equation}
    g_{\rm t}(R) = \frac{\gamma_{\rm t}(R)}{1 - \kappa(R)}\, .
\end{equation}

Above, $\gamma_{\rm t}(R) \equiv \Delta \Sigma(R) / \Sigma_{\rm crit}$ and $\kappa(R) \equiv \Sigma(R) / \Sigma_{\rm crit}$, where the critical surface density is defined as

\begin{equation}
    \Sigma_{\rm crit} \equiv \frac{{\rm c}^2}{4 \uppi \mathrm{G}}\frac{D_{\rm s}}{D_{\rm l} D_{\rm ls}}\, ,
\end{equation}

\noindent where $D_{\rm l}$, $D_{\rm ls}$, and $D_{\rm s}$ are respectively the angular diameter distances between lens-observer, lens-source, and source-observer. In the case of weak lensing, $\kappa \ll 1$ so that $g_{\rm t} (R) \simeq \gamma_{\rm t} (R) \propto \Delta \Sigma$.
Consequently, weak lensing allows reconstruction of the mass and density profiles of halos independently of the dynamical state of the system. However, due to the subtle nature of weak lensing distortions, modern observational surveys can only robustly detect the individual lensing signal associated with massive halos at the cluster level (e.g.,~\citealt{Sartoris_2016}). Therefore, we further restricted our analysis by performing a mass cut at $M_{200} \simeq 10^{13}\ \text{M}_{\odot}$, which reduced our sample to $\sim 370$ halos.

\begin{figure}[ht!]
    \centering
    \includegraphics[width=\hsize]{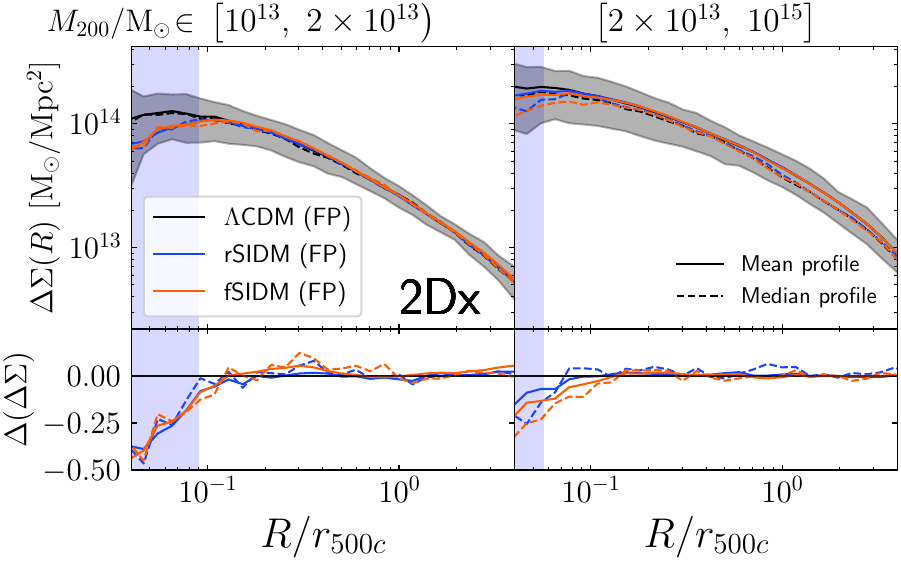}
    \caption{Stacked DM surface density excess profiles for the FP simulations, computed along the $x$-direction. The color code is the same as in Fig.~\ref{fig:mass_profiles}.  The bottom panel shows the relative differences between each model for both mean and median profiles, where $\Delta (\Delta \Sigma) \equiv (\Delta \Sigma - \Delta \Sigma_{\rm CDM}) / \Delta \Sigma_{\rm CDM}$.}
    \label{fig:dsigma_profiles}
\end{figure}

\begin{figure}[ht!]
    \centering
    \includegraphics[width=\hsize]{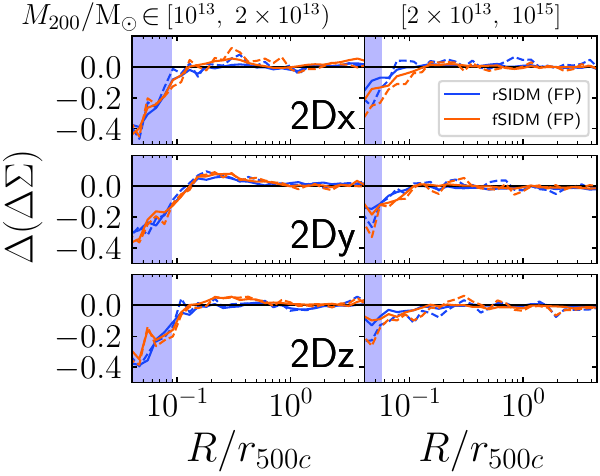}
    \caption{Dark matter $\Delta \Sigma$ stacked profiles relative differences compared to CDM along the $x$, $y$, and $z$ directions. The color code and line styles are the same as in Fig.~\ref{fig:mass_profiles}.}
    \label{fig:dsigma_profiles_proj}
\end{figure}

Figure~\ref{fig:dsigma_profiles} shows the FP stacked excess surface density profiles along the $x$-direction (similar results hold for the other projections; see Fig.~\ref{fig:dsigma_profiles_proj}) computed in two mass intervals, $M_{200} / \text{M}_\odot \in \left[ 10^{13},\ 2 \times 10^{13} \right)$ and $M_{200} / \text{M}_\odot \in \left[ 2 \times 10^{13},\ 10^{15} \right]$, chosen to both contain roughly the same number of halos. Interestingly, unlike the projected mass and density (e.g., Fig.~\ref{fig:mass_profiles}), the surface density excess profiles are characterized by a greater SIDM suppression at small radii, reaching deviations of $\sim 10\%$ (above the convergence threshold) and potentially up to $\sim 40\%$ (below the threshold) in the lower mass bin and $\sim 20\%$ above the threshold for higher masses, implying that weak lensing shear mass reconstructions might potentially serve as a robust probe of SIDM.
Finally, the mass bump visible in the 3D and 2D mass profiles hardly appears in the rSIDM lensing profiles at these mass scales. 

Being a 2D quantity, the surface density excess is particularly affected by systematic projection effects, which can bias the profile estimate. For example, the impact associated with mis-centering of the estimated halo center from the true potential minimum~\citep{Martin_2023} and overestimation of the central mass due to spurious large-scale structure along the line of sight~\citep{EP-Ragagnin} may severely impact the deviations of SIDM at the $\sim 10\%$ level.
Moreover, other sources of systematic uncertainty come in the form of halo triaxiality and orientation bias (e.g.,~\citealt{Becker_2011, Meneghetti_2014, Giocoli_2024}) -- which may play a significant part in this case, as SIDM halos are expected to be closer to being spherically symmetric -- influencing the relative differences with CDM depending on the direction of the line of sight. In Fig.~\ref{fig:dsigma_profiles_proj} we show the relative differences between $\Delta \Sigma$ computed along all three Cartesian axes. While the lower-mass SIDM profiles seem to exhibit a slightly weaker central suppression and stronger density bump in the 2Dy case, overall we did not observe any particular differences between the three lines of sight, suggesting that the stacked profiles may wash out any eventual projection biases resulting from individual halo profiles. Similarly, the FP median c-M relation studied in Sect.~\ref{sec:c_m_relation} also indicates a deviation of 2Dy from the behavior of the 2Dx and 2Dz projections (see Fig.~\ref{fig:cM_median_diff}); however, it does not significantly influence our results.

Finally, though our aim was to provide a simple estimate of the weak lensing induced by SIDM, we have to note that in the denser cores of halos (where $\kappa \gg 1$), we expected strong lensing effects to gain dominance, meaning that the surface density excess alone would no longer accurately reflect the behavior of lensing.
However, an in-depth analysis of strong lensing would require the use of more complex ray-tracing simulations to accurately follow the path of lensed photons, which is beyond the scope of this work.

\subsection{Comparison with observational profiles} \label{sec:obs_cluster}
We compared our theoretical profiles with the reconstructed mass distribution of real galaxy clusters. In an observational context, galaxy clusters offer a reliable way to test nonstandard models, providing a robust probe of the halo DM distribution through weak lensing and kinematic galaxy measurements. Therefore, we studied the 3D cumulative DM mass profile of the MACS~J1206 and Abell~S1063 galaxy clusters, which respectively have a total mass of $M_{\rm tot}^\mathrm{MACS} \simeq 1.56 \times 10^{15}\ \text{M}_\odot$ \citep{Biviano_2023} and $M_{\rm tot}^\mathrm{Abell} \simeq 3.08 \times 10^{15}\ \text{M}_\odot$ \citep{Sartoris_2020} and were observed at redshift $z \simeq 0.4$ using galaxy data from the CLASH survey. The choice of these clusters is based on both the availability of high precision measurements for their galaxy tracers -- more than $\sim 500$ galaxies within $r_{200c}$ with accurate spectroscopic redshift plus high-quality kinematic information on the brightest cluster galaxy (e.g., \citealt{Mercurio_2021}), provided by the VIMOS and MUSE spectrographs at the Very Large Telescope -- and in the case of MACS J1206, on its regular dynamical structure (i.e.,\ approximately spherically symmetric and dynamically relaxed, e.g.,~\citealt[][]{Girardi2015,Pizzuti_2022}), which avoids complications due to cluster relaxation processes. On the simulation side, we limited our analysis to our six largest DIANOGA-SIDM halos, whose masses are reported in Table~\ref{tab:DIANOGA_regions}.

We used the multicomponent mass profile of MACS~J1206 derived by \cite{Biviano_2023} from the kinematics of member galaxies and the stellar velocity dispersion of the central brightest galaxy, using the \MGMAMPOSSt code~\citep{Pizzuti_2021}. They found an internal logarithmic slope of the DM distribution, parametrized by a generalized NFW model~\citep{Wyithe_2001}, of $\gamma_{\rm DM} = 0.7^{+0.2}_{-0.1}$ at $1\,\sigma$. \cite{Sartoris_2020} performed a mass reconstruction of Abell S1063. The DM profile was also modeled with a generalized NFW, with a slope of $\gamma_{\rm DM} = 0.99 \pm 0.04$ (i.e., in agreement with CDM predictions).

\begin{figure}[ht!] \label{fig:cluster_observ}
    \centering
    \includegraphics[width=0.9\hsize]{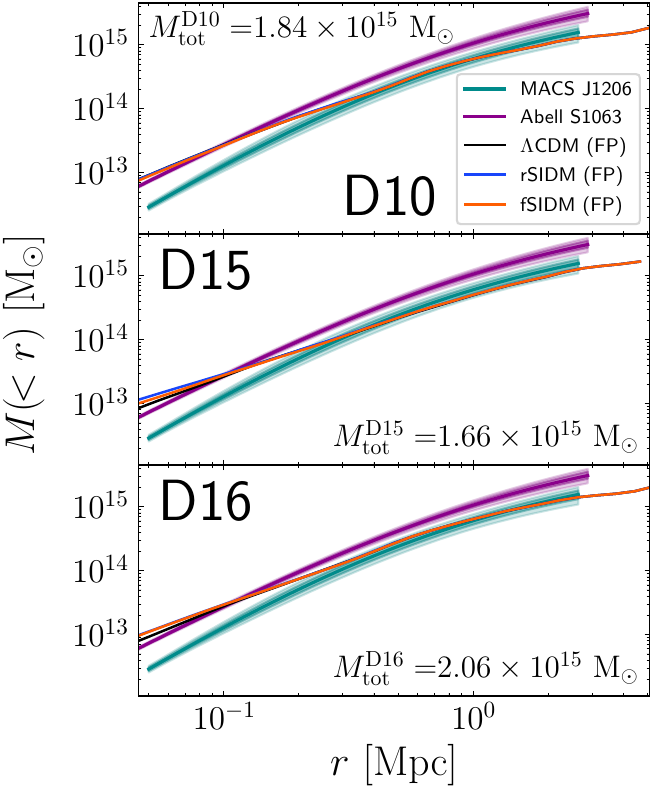}
    \caption{Comparison of the DM mass profiles of our simulated cluster-sized halos belonging to the D10 (\textit{top}), D15 (\textit{middle}), and D16 (\textit{bottom}) regions and the real galaxy clusters MACS~J1206 (the blue line;~\citealt{Biviano_2023}) and Abell~S1063 (the purple line;~\citealt{Sartoris_2020}). The comparison is shown for all DM models using the FP simulations.}
    \label{fig:mass_observ}
\end{figure}

In the following we only show a comparison with our FP simulated halos that possess a total mass similar to the observed clusters. These halos are the most massive halos associated with the galaxy clusters in the regions D10, D15, and D16, with $M_{\rm tot}^{\rm D10} \simeq 1.84 \times 10^{15}\ \text{M}_\odot$, $M_{\rm tot}^{\rm D15} \simeq 1.66 \times 10^{15}\ \text{M}_\odot$, and $M_{\rm tot}^{\rm D16} \simeq 2.06 \times 10^{15}\ \text{M}_\odot$, respectively (see Fig.~\ref{fig:mass_observ}). In the plots we show the DM mass profiles of the simulated halos with different colored lines. The MACS~J1206 and Abell~S1063 DM mass profiles -- along with their $68\%$ and $95\%$ credible intervals -- are respectively shown with solid cyan and purple curves. In this case, the region of convergence for our simulated profiles (see Sect.~\ref{sec:stacked_profiles}) extends below our studied radial range, so it is not shown.

Figure~\ref{fig:mass_observ} shows that all of our DM models agree fairly well with the mass profile shape of MACS~J1206 in the halo outskirts. However, for radii $r \lesssim 0.2$ -- $0.3\ \text{Mpc}$, all simulations (including CDM) present a significant deviation from both of the observed clusters, exhibiting much cuspier inner profiles.  
We note, however, that due to the lack of a statistically significant sample of both simulated and observed massive halos, for our comparison we considered individual halo profiles instead of stacked quantities, which in general is not representative of the typical halo population. Despite the halos we considered having similar total masses, their diverse assembly histories could result in very different inner halo structures -- especially when considering the presence of baryonic interactions and complex feedback mechanisms -- for both the simulated and observed halos. In particular, we report in Appendix~\ref{app:gas_star_profiles} the mass profiles for the gas and stellar components of our simulated and observed clusters. It is clear that the central gas and stellar components in the simulations dominate over those of MACS~J1206 and Abell~S1063 for all DM models considered, implying a greater impact of adiabatic contraction inside such clusters and thus leading to more pronounced differences between simulated and observed DM profiles.
Moreover, the presence of observational projection systematics can significantly bias the inferred profiles. In this regard, we expect a more detailed inference of the MACS~J1206 mass profile in a future publication that more robustly incorporates the effects of both weak and strong lensing alongside galaxy kinematics (Pizzuti et al., in prep.).

\begin{figure}[ht!]
    \centering
    \includegraphics[width=\hsize]{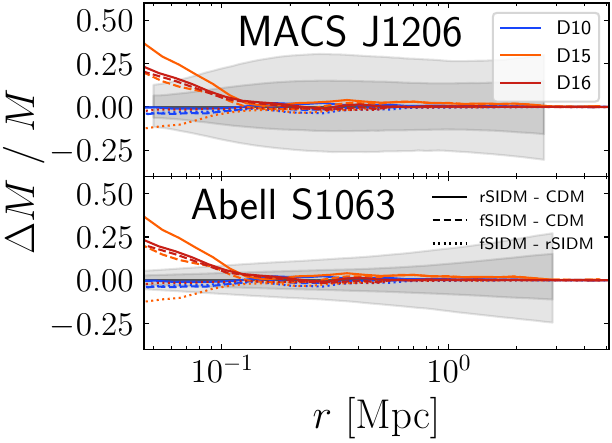}
    \caption{Relative differences between rSIDM-CDM (solid lines), fSIDM-CDM (dashed lines), and fSIDM-rSIDM (dotted lines) of the D10, D15, and D16 region DM halo mass profiles compared to the MACS~J1206 (\textit{top}) and Abell~S1063 (\textit{bottom}) uncertainties (gray shaded regions).
    }
    \label{fig:obs_diffs}
\end{figure}

Consequently, the high variability of individual halo profiles cannot allow for a conclusive, absolute comparison between our profiles. However, given the focus of this work on assessing the observational viability of halo profiles as probes of SIDM, we provide a comparison between observational uncertainties and SIDM relative deviations. In this regard, it is interesting to remark how in our few cluster-sized halos, the SIDM profiles tend to be consistently cuspier than CDM in the very inner core (see, e.g., Fig.~\ref{fig:obs_diffs}), which is unlike the behavior observed in the lower-mass halos -- even when considering the higher mass interval (see, e.g., Fig.~\ref{fig:mass_profiles}). Because such behavior does not seem to be present in the DMO case, it must be attributed to the presence of baryons. While SIDM-induced core-collapse can potentially produce cuspy inner profiles in very massive halos, the cross sections needed for this process to occur on a timescale shorter than the Hubble time are much larger than those considered in this work~\citep{Zhong_2023}.
Only if the baryons are dominant enough in the center to erase the positive velocity dispersion gradient of the DM does the halo not undergo a core formation phase, and instead it begins to collapse immediately \citep[e.g.,][]{Rose_2023, Zeng_2026}. However, in our simulations, the inner velocity dispersion gradient remains positive \citep{Ragagnin_2024}, indicating core formation. This suggests that the evolution of a SIDM halo, including various baryonic processes and its cosmological assembly history, substantially complicates its evolution \citep[e.g.,][]{Silverman_2026}.
\footnote{Other interesting cases that may or may not be related to our findings have been reported by \cite{Robles_2019}, as shown in their Fig.~1, and by \cite{vanDenBosch2026}, as shown in their Fig.~6.}
The exact origin of this behavior remains unclear, but considering its presence in other simulations and the small convergence radius for our massive halos possibly disconnects it from being a simple numerical artifact or resolution-based effect. 

In the case that this SIDM enhancement is connected to a real physical phenomenon, it would potentially change our understanding of how DM self-interactions shape the cores of galaxy clusters and how they may be utilized as probes of SIDM effects. Nevertheless, this enhancement could be used to constrain SIDM. Indeed, Fig.\ref{fig:obs_diffs} shows the relative differences between our simulated profiles compared to the relative uncertainties of the MACS~J1206 (top) and Abell~S1063 (bottom) profiles. As these uncertainties decrease significantly at small radii ($r \lesssim 0.1$ Mpc), it is clear that both rSIDM and fSIDM (solid and dashed lines) may produce potentially detectable imprints on the mass profiles of our most massive halos.
In this regard, the cores of very massive clusters might offer the opportunity for precise constraints on SIDM models.

\subsection{Concentration-mass relation} \label{sec:c_m_relation}
We conclude by briefly exploring the relation between halo concentration $c_{200}$ and virial mass $M_{200}$. The concentration parameter is a measure of the compactness of a DM halo, acting as a fundamental structural parameter due to its close connection to the halo formation and evolution history (e.g.,~\citealt{Wechsler_2002}). Assuming an NFW profile, the concentration represents how much larger the virial radius is when compared to the scale radius:

\begin{equation}
    c_{200} \equiv \frac{r_{200c}}{r_{\rm s}}\, .
\end{equation}

In general, in hierarchical structure formation, low-mass \lcdm halos are expected to be denser and more concentrated than more massive ones, due to their early formation. Consequently, $c_{200}$ exhibits a decreasing trend with halo mass~\citep[e.g.,][]{Child_2018}. In this section we assess whether SIDM models significantly affect this behavior. 

To this end, we computed the concentration parameter obtained from the NFW mass fits, both 3D and projected. 
Although perhaps too simple to accurately model SIDM imprints in the inner halo, the NFW model is still a quite popular choice to extract halo concentrations in an observational context, so we report the NFW results to provide a closer match to observational techniques.
To obtain a statistically robust estimate of the concentration trend with mass, we computed the median $c_{200}$ across three $\log{M_{200}}$ bins, with each bin containing roughly the same number of halos.

\begin{figure*}[ht!]
    \centering
    \includegraphics[width=0.9\hsize]{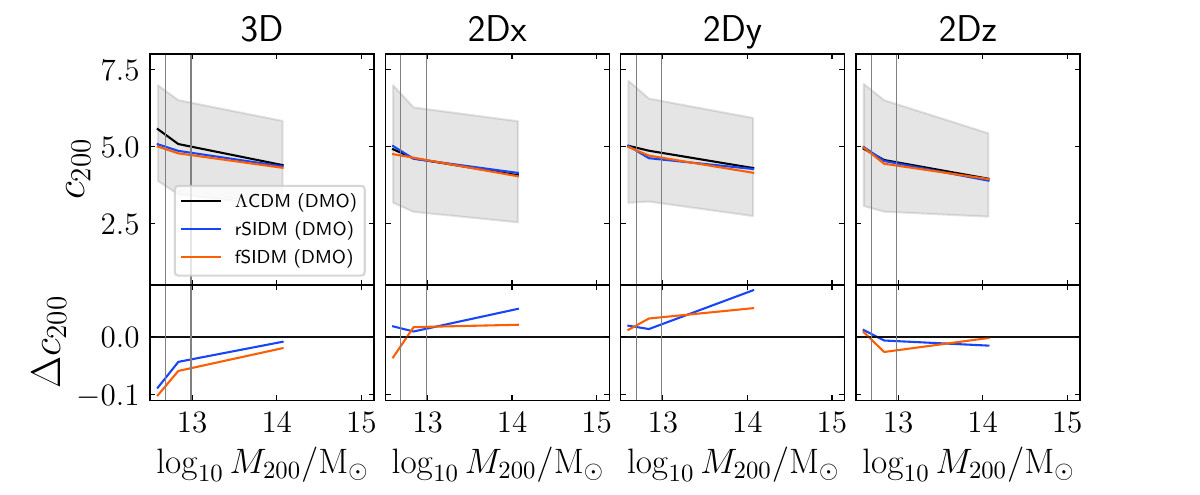}
    \includegraphics[width=0.9\hsize]{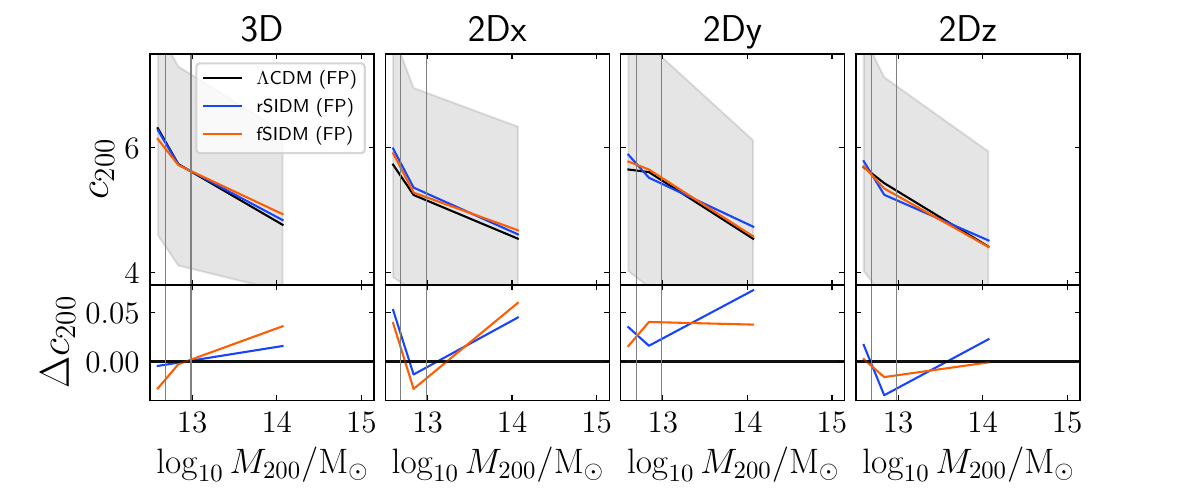}
    \caption{Median concentrations for all DMO (\textit{top}) and FP (\textit{bottom}) simulations for all 3D and 2D cases computed across three logarithmically spaced mass bins selected to each be populated by roughly the same number of halos. The edges of the bins are indicated by the vertical gray lines. The gray shaded region delimits the $68\%$ scatter for the \lcdm model. Relative differences with CDM are shown in the bottom panels, where $\Delta c_{200} \equiv (c_{200} - c_{200, \rm CDM}) / c_{200, \rm CDM}$. All halo concentrations and virial masses were extracted from NFW mass fits (see Sect.~\ref{sec:mass_fits}).}
    \label{fig:cM_median_diff}
\end{figure*}

The resulting c-M relations and their median trend are shown in Appendix~\ref{app:cM_relations}, from which it is clear that all FP simulations are characterized by a greater $c_{200}$ intrinsic scatter in the low-mass end, especially for the 2D cases. The comparison of median concentrations between all models is shown in Fig.~\ref{fig:cM_median_diff}. In the 3D DMO case, both SIDM models tend to systematically suppress the concentration of $M_{200} < 10^{13}\ \mathrm{M_\odot}$ halos at the $\sim 10\%$ level, with a stronger effect in fSIDM.
The addition of baryons tends to significantly increase the average concentration at all mass scales, resulting in 3D concentrations in the higher-mass bin that are actually greater than CDM on average but with deviations of only $\lesssim 5\%$.

Finally, while the 2D concentrations are also enhanced in the presence of baryons, both DMO and FP cases only exhibit SIDM differences of $\lesssim 5\%$ along all projections. 
Observationally, this may suggest that computation of the NFW c-M relation from current weak lensing measurements might not provide a fruitful way of distinguishing between SIDM models when compared to direct analysis of the SIDM surface density excess profiles, which present a much clearer deviation from CDM (see Fig.~\ref{fig:dsigma_profiles}). Importantly, we highlight that the low flexibility of the simple NFW model does not allow the cored internal profiles produced by SIDM to be fully captured, leading to a biased estimation of the halo concentrations. Therefore, these results should also encourage future weak lensing c-M relation computations to utilize halo models with a higher number of free parameters or to find other suitable parameterizations (see, e.g.,~\citealt{ODonnell_2026}) in order to better reconstruct the internal structure of SIDM halos.

\section{Summary and conclusions}
\label{sec:conc}
In this work, we have conducted an analysis of DM halo profiles of groups and clusters of galaxies in order to investigate the complex interplay between baryons and DM in the SIDM framework with the aim of informing future weak lensing and galaxy kinematics observational analyses in which halo properties will serve as sensitive probes of SIDM. In particular, we extracted halo mass and density profiles from the DIANOGA-SIDM simulations -- which feature rare and frequent collision-type SIDM models -- both for the DMO case and for the case where full baryonic physics is included (FP). We performed stacking and fitting to robustly explore the inner regions of halos ($r \lesssim r_{500c}$) where SIDM is expected to play a prominent role. The main results of our study are summarized as follows:

\begin{itemize}
    \item In lower-mass halos, SIDM effectively produces more cored internal 3D profiles than CDM, with relative differences of $\sim 20\%$ in the DMO case. The FP case results in smaller deviations since baryons counteract the effects of SIDM. Both DMO and FP SIDM profiles are also characterized by a $\sim 5\%$ mass excess at intermediate radii -- which is especially evident in the density profiles -- as a consequence of heat transfer from collisions. Conversely, the most massive SIDM halos are more affected by baryons, leading to cuspier inner profiles than CDM and a much less pronounced mass excess at larger radii.
    \item While projecting the mass and density profiles along the line of sight reduces the effects of SIDM at all halo masses -- leading to cuspier profiles than the 3D case -- the weak lensing shear (surface density excess) profiles still reach deviations of up to $\sim 20\%$ above the resolution limit for $M_{200} \sim 10^{13}\ \text{M}_\odot$ halos along all lines of sight considered.
    \item We compared the DM profiles of the most massive halos in our simulations to those inferred from kinematical analysis of the massive clusters MACS~J1206 and Abell~S1063 ($M_{\rm tot} \sim 10^{15}\ \text{M}_\odot$; see~\citealt{Sartoris_2020, Biviano_2023}). We find that while none of our individual halos match perfectly with the observed profiles, the observational uncertainty in the cluster core might be small enough to allow SIDM models to be distinguished.
    \item The median NFW concentration remains only slightly affected by SIDM at all mass scales, with deviations at the $\lesssim 5\%$ level in the projected FP case.
\end{itemize}

It is striking that simulated cluster cores exhibit significant SIDM enhancement and that the observational uncertainty decreases enough to allow, in principle, for significant detection using galaxy kinematic measurements (see Fig.~\ref{fig:obs_diffs}). Therefore it may also be fruitful to further explore the role played by SIDM in the cores of very massive clusters in order to gain a more complete understanding of the SIDM central mass excess and its role in constraining SIDM physics. In this case, due to the lack of statistics of our massive halo sample, we were not able to perform an effective stacking analysis. Thus future work may address this issue by running sufficiently larger-scale simulations. Furthermore, our study suggests that weak lensing inference of the surface density excess profiles could prove particularly useful for detecting SIDM effects in $M_{200} \sim 10^{13}\ \text{M}_\odot$ clusters, although further modeling of observational systematics is required for a more realistic detection analysis.

Finally, we note that given the relatively limited number of DM particles, there is a certain noise in the evolution time of our SIDM halos~\citep[see, e.g.,][]{Mace_2024} \footnote{We note, however, that the problem of the realization noise when sampling an isolated low-resolution halo with \textsc{SpherIC} \citep{Garrison-Kimmel_2013} as found by \cite{Mace_2024} \citep[see also Appendix~C in][]{Schmidt_2026} does not apply to our cosmological simulations.} that may influence our results.
In this regard, a detailed study of the numerical errors in the gravothermal evolution of SIDM halos with our code was undertaken by \citet{Fischer_2024b, Fischer_2025}. Employing future simulations with a higher number of particles will therefore be crucial to providing more accurate results.

Overall, despite SIDM imprints on halo mass and density profiles being washed out from both baryons and 2D projection -- thus reducing the observational viability of such probes -- we have also showed that there are certain observationally promising radial scales that may potentially act as robust probes of such SIDM effects through weak lensing (such as in the internal regions of group-sized halos; see Fig.~\ref{fig:dsigma_profiles_proj}) and galaxy-traced mass reconstructions (such as in the cores of very massive clusters; see Fig.~\ref{fig:obs_diffs}).
While our investigation primarily focused on the shapes of halo mass and density profiles, we aim to further expand our analysis of the DIANOGA-SIDM simulations in a follow-up paper, where we will study the velocity dispersion profiles of halos, focusing particularly on the shapes of DM particle orbits, in order to assess if the halo velocity anisotropy could potentially represent a viable probe for the detection of SIDM. Furthermore, we will also explore in detail the role played by SIDM within merging systems, evaluating whether their matter distribution may reveal significant signatures of DM self-interactions.

\begin{acknowledgements}
The authors acknowledge A. Benitez-Llambay and G. Bordogna for useful suggestions and tests to improve the analysis.
MSF gratefully acknowledges the support of the Alexander von Humboldt Foundation through a Feodor Lynen Research Fellowship. 
We acknowledge usage of INAF IT framework~\citep{Bertocco_2020,Taffoni_2020}.
We used \texttt{python} packages \texttt{colossus}~\citep{Diemer_2018}
and \texttt{numpy}~\citep{Harris_2020}.
AR acknowledges EuroHPC Joint Undertaking for awarding the project ID EHPC-REG-2024R01-029 access to Leonardo at CINECA, Italy; 
 and for awarding us access to MareNostrum5 as BSC, Spain (ID EHPC-BEN-2025B11-033) and LUMI at CSC, Finland (ID EHPC-DEV-2026D02-240).
AR acknowledges ISCRA for awarding this project access to the LEONARDO supercomputer, owned by the EuroHPC Joint Undertaking, hosted by CINECA, Italy (HP10BUFI59).
LP acknowledges support by the Italian Ministry for Research and University (MUR) under 457 Grant ``Progetto Dipartimenti di Eccellenza 2023-2027'' (BiCoQ).

\end{acknowledgements}

\bibliographystyle{bibtex/aa}
\bibliography{bibliography.bib}

@inproceedings{Rosati2014,
  author    = {Rosati, Piero and Balestra, Italo and Grillo, Claudio and Mercurio, Amata and Nonino, Mario and Biviano, Andrea and Girardi, Marisa and Vanzella, Eros and the CLASH-VLT Team},
  title     = {CLASH-VLT: A VIMOS Large Programme to Map the Dark Matter Mass Distribution in Galaxy Clusters and Probe Distant Lensed Galaxies},
  booktitle = {Astronomical Science},
  year      = {2014},
  pages      = {40--43},
}

@article{Umetsu25,
doi = {10.3847/1538-4357/aded91},
url = {https://doi.org/10.3847/1538-4357/aded91},
year = {2025},
month = {aug},
publisher = {The American Astronomical Society},
volume = {990},
number = {1},
pages = {70},
author = {Umetsu, Keiichi and Pizzardo, Michele and Diaferio, Antonaldo and Geller, Margaret J.},
title = {Cluster Lensing Mass Inversion (CLUMI+): Combining Dynamics and Weak Lensing around Galaxy Clusters},
journal = {\apj}
}

@ARTICLE{Vegetti_2024,
       author = {{Vegetti}, S. and {Birrer}, S. and {Despali}, G. and {Fassnacht}, C.~D. and {Gilman}, D. and {Hezaveh}, Y. and {Perreault Levasseur}, L. and {McKean}, J.~P. and {Powell}, D.~M. and {O'Riordan}, C.~M. and {Vernardos}, G.},
        title = "{Strong Gravitational Lensing as a Probe of Dark Matter}",
      journal = {\ssr},
         year = 2024,
        month = aug,
       volume = {220},
       number = {5},
          eid = {58},
        pages = {58},
          doi = {10.1007/s11214-024-01087-w},
archivePrefix = {arXiv},
       eprint = {2306.11781},
 primaryClass = {astro-ph.CO},
       adsurl = {https://ui.adsabs.harvard.edu/abs/2024SSRv..220...58V}
}

@article{Biviano_2023,
doi = {10.3847/1538-4357/acf832},
url = {https://doi.org/10.3847/1538-4357/acf832},
year = {2023},
month = {nov},
publisher = {The American Astronomical Society},
volume = {958},
number = {2},
pages = {148},
author = {Biviano, Andrea and Pizzuti, Lorenzo and Mercurio, Amata and Sartoris, Barbara and Rosati, Piero and Ettori, Stefano and Girardi, Marisa and Grillo, Claudio and Caminha, Gabriel B. and Nonino, Mario},
title = {CLASH-VLT: The Inner Slope of the MACS J1206.2-0847 Dark Matter Density Profile},
journal = {\apj}
}

@article{Pizzuti_2021,
    author = {Pizzuti, Lorenzo and Saltas, Ippocratis D and Amendola, Luca},
    title = {mg-mamposst: a code to test modifications of gravity with internal kinematics and lensing analyses of galaxy clusters},
    journal = {\mnras},
    volume = {506},
    number = {1},
    pages = {595-612},
    year = {2021},
    month = {09},
    issn = {0035-8711},
    doi = {10.1093/mnras/stab1727},
    url = {https://doi.org/10.1093/mnras/stab1727},
    eprint = {https://academic.oup.com/mnras/article-pdf/506/1/595/38884757/stab1727.pdf},
}

@article{Sirks_2024,
    author = {Sirks, Ellen L and Harvey, David and Massey, Richard and Oman, Kyle A and Robertson, Andrew and Frenk, Carlos and Everett, Spencer and Gill, Ajay S and Lagattuta, David and McCleary, Jacqueline},
    title = {Hydrodynamical simulations of merging galaxy clusters: giant dark matter particle colliders, powered by gravity},
    journal = {\mnras},
    volume = {530},
    number = {3},
    pages = {3160-3170},
    year = {2024},
    month = {05},
    issn = {0035-8711},
    doi = {10.1093/mnras/stae1012},
    url = {https://doi.org/10.1093/mnras/stae1012},
    eprint = {https://academic.oup.com/mnras/article-pdf/530/3/3160/57368167/stae1012.pdf},
}

@article{Sirks_2022,
    author = {Sirks, Ellen L and Oman, Kyle A and Robertson, Andrew and Massey, Richard and Frenk, Carlos},
    title = {The effects of self-interacting dark matter on the stripping of galaxies that fall into clusters},
    journal = {\mnras},
    volume = {511},
    number = {4},
    pages = {5927-5935},
    year = {2022},
    month = {04},
    issn = {0035-8711},
    doi = {10.1093/mnras/stac406},
    url = {https://doi.org/10.1093/mnras/stac406},
    eprint = {https://academic.oup.com/mnras/article-pdf/511/4/5927/42793992/stac406.pdf},
}

@article{Sabarish_2024,
    author = {Sabarish, V M and Brüggen, Marcus and Schmidt-Hoberg, Kai and Fischer, Moritz S and Kahlhoefer, Felix},
    title = {Simulations of galaxy cluster mergers with velocity-dependent, rare, and frequent self-interactions},
    journal = {\mnras},
    volume = {529},
    number = {3},
    pages = {2032-2046},
    year = {2024},
    month = {04},
    issn = {0035-8711},
    doi = {10.1093/mnras/stae664},
    url = {https://doi.org/10.1093/mnras/stae664},
    eprint = {https://academic.oup.com/mnras/article-pdf/529/3/2032/57001559/stae664.pdf},
}

@article{Tokayer_2024,
doi = {10.3847/1538-4357/ad51fd},
url = {https://doi.org/10.3847/1538-4357/ad51fd},
year = {2024},
month = {jul},
publisher = {The American Astronomical Society},
volume = {970},
number = {2},
pages = {143},
author = {Tokayer, Yarone M. and Dutra, Isaque and Natarajan, Priyamvada and Mahler, Guillaume and Jauzac, Mathilde and Meneghetti, Massimo},
title = {The Galaxy–Galaxy Strong Lensing Cross Section and the Internal Distribution of Matter in ΛCDM Substructure},
journal = {\apj}
}

@article{Chiang_2026,
doi = {10.3847/1538-4357/ae23c1},
url = {https://doi.org/10.3847/1538-4357/ae23c1},
year = {2026},
month = {jan},
publisher = {The American Astronomical Society},
volume = {997},
number = {1},
pages = {106},
author = {Chiang, Barry T. and Dutra, Isaque and Natarajan, Priyamvada},
title = {Constraining the Nature of Dark Matter from Tidal Radii of Cluster Galaxy Subhalos},
journal = {\apj}
}

@article{Natarajan_2026,
doi = {10.3847/2041-8213/ae53ea},
url = {https://doi.org/10.3847/2041-8213/ae53ea},
year = {2026},
month = {apr},
publisher = {The American Astronomical Society},
volume = {1001},
number = {1},
pages = {L12},
author = {Natarajan, Priyamvada and Chiang, Barry T. and Dutra, Isaque},
title = {New Cold Dark Matter Crisis Revealed by Multiscale Cluster Lensing},
journal = {\apj}
}

@article{Dutra_2025,
doi = {10.3847/1538-4357/ad9b09},
url = {https://doi.org/10.3847/1538-4357/ad9b09},
year = {2024},
month = {dec},
publisher = {The American Astronomical Society},
volume = {978},
number = {1},
pages = {38},
author = {Dutra, Isaque and Natarajan, Priyamvada and Gilman, Daniel},
title = {Self-interacting Dark Matter, Core Collapse, and the Galaxy–Galaxy Strong-lensing Discrepancy},
journal = {\apj}
}

@ARTICLE{Ragagnin_2022,
       author = {{Ragagnin}, Antonio and {Meneghetti}, Massimo and {Bassini}, Luigi and {Ragone-Figueroa}, Cinthia and {Granato}, Gian Luigi and {Despali}, Giulia and {Giocoli}, Carlo and {Granata}, Giovanni and {Moscardini}, Lauro and {Bergamini}, Pietro and {Rasia}, Elena and {Valentini}, Milena and {Borgani}, Stefano and {Calura}, Francesco and {Dolag}, Klaus and {Grillo}, Claudio and {Mercurio}, Amata and {Murante}, Giuseppe and {Natarajan}, Priyamvada and {Rosati}, Piero and {Taffoni}, Giuliano and {Tornatore}, Luca and {Tortorelli}, Luca},
        title = "{Galaxies in the central regions of simulated galaxy clusters}",
      journal = {\aap},
         year = 2022,
        month = sep,
       volume = {665},
          eid = {A16},
        pages = {A16},
          doi = {10.1051/0004-6361/202243651},
archivePrefix = {arXiv},
       eprint = {2204.09067},
 primaryClass = {astro-ph.CO},
       adsurl = {https://ui.adsabs.harvard.edu/abs/2022A&A...665A..16R}
}

@ARTICLE{Meneghetti_2023,
       author = {{Meneghetti}, Massimo and {Cui}, Weiguang and {Rasia}, Elena and {Yepes}, Gustavo and {Acebron}, Ana and {Angora}, Giuseppe and {Bergamini}, Pietro and {Borgani}, Stefano and {Calura}, Francesco and {Despali}, Giulia and {Giocoli}, Carlo and {Granata}, Giovanni and {Grillo}, Claudio and {Knebe}, Alexander and {Macci{\`o}}, Andrea V. and {Mercurio}, Amata and {Moscardini}, Lauro and {Natarajan}, Priyamvada and {Ragagnin}, Antonio and {Rosati}, Piero and {Vanzella}, Eros},
        title = "{A persistent excess of galaxy-galaxy strong lensing observed in galaxy clusters}",
      journal = {\aap},
         year = 2023,
        month = oct,
       volume = {678},
          eid = {L2},
        pages = {L2},
          doi = {10.1051/0004-6361/202346975},
archivePrefix = {arXiv},
       eprint = {2309.05799},
 primaryClass = {astro-ph.CO},
       adsurl = {https://ui.adsabs.harvard.edu/abs/2023A&A...678L...2M}
}

@ARTICLE{Meneghetti_2022,
       author = {{Meneghetti}, Massimo and {Ragagnin}, Antonio and {Borgani}, Stefano and {Calura}, Francesco and {Despali}, Giulia and {Giocoli}, Carlo and {Granato}, Gian Luigi and {Grillo}, Claudio and {Moscardini}, Lauro and {Rasia}, Elena and {Rosati}, Piero and {Angora}, Giuseppe and {Bassini}, Luigi and {Bergamini}, Pietro and {Caminha}, Gabriel B. and {Granata}, Giovanni and {Mercurio}, Amata and {Metcalf}, Robert Benton and {Natarajan}, Priyamvada and {Nonino}, Mario and {Pignataro}, Giada Venusta and {Ragone-Figueroa}, Cinthia and {Vanzella}, Eros and {Acebron}, Ana and {Dolag}, Klaus and {Murante}, Giuseppe and {Taffoni}, Giuliano and {Tornatore}, Luca and {Tortorelli}, Luca and {Valentini}, Milena},
        title = "{The probability of galaxy-galaxy strong lensing events in hydrodynamical simulations of galaxy clusters}",
      journal = {\aap},
         year = 2022,
        month = dec,
       volume = {668},
          eid = {A188},
        pages = {A188},
          doi = {10.1051/0004-6361/202243779},
archivePrefix = {arXiv},
       eprint = {2204.09065},
 primaryClass = {astro-ph.CO},
       adsurl = {https://ui.adsabs.harvard.edu/abs/2022A&A...668A.188M}
}

@ARTICLE{Meneghetti_2020,
       author = {{Meneghetti}, Massimo and {Davoli}, Guido and {Bergamini}, Pietro and {Rosati}, Piero and {Natarajan}, Priyamvada and {Giocoli}, Carlo and {Caminha}, Gabriel B. and {Metcalf}, R. Benton and {Rasia}, Elena and {Borgani}, Stefano and {Calura}, Francesco and {Grillo}, Claudio and {Mercurio}, Amata and {Vanzella}, Eros},
        title = "{An excess of small-scale gravitational lenses observed in galaxy clusters}",
      journal = {Sci},
         year = 2020,
        month = sep,
       volume = {369},
       number = {6509},
        pages = {1347-1351},
          doi = {10.1126/science.aax5164},
archivePrefix = {arXiv},
       eprint = {2009.04471},
 primaryClass = {astro-ph.GA},
       adsurl = {https://ui.adsabs.harvard.edu/abs/2020Sci...369.1347M}
}

@ARTICLE{Correa_2021,
       author = {{Correa}, Camila A.},
        title = "{Constraining velocity-dependent self-interacting dark matter with the Milky Way's dwarf spheroidal galaxies}",
      journal = {\mnras},
         year = 2021,
        month = may,
       volume = {503},
       number = {1},
        pages = {920-937},
          doi = {10.1093/mnras/stab506},
archivePrefix = {arXiv},
       eprint = {2007.02958},
 primaryClass = {astro-ph.GA},
       adsurl = {https://ui.adsabs.harvard.edu/abs/2021MNRAS.503..920C}
}

@ARTICLE{Correa_2025,
       author = {{Correa}, Camila A. and {Schaller}, Matthieu and {Schaye}, Joop and {Ploeckinger}, Sylvia and {Borrow}, Josh and {Bah{\'e}}, Yannick},
        title = "{TangoSIDM Project: is the stellar mass Tully-Fisher relation consistent with SIDM?}",
      journal = {\mnras},
         year = 2025,
        month = feb,
       volume = {536},
       number = {4},
        pages = {3338-3356},
          doi = {10.1093/mnras/stae2811},
archivePrefix = {arXiv},
       eprint = {2403.09186},
 primaryClass = {astro-ph.CO},
       adsurl = {https://ui.adsabs.harvard.edu/abs/2025MNRAS.536.3338C}
}

@ARTICLE{Shah_2024,
       author = {{Shah}, Neev and {Adhikari}, Susmita},
        title = "{The abundance of core-collapsed subhaloes in SIDM: insights from structure formation in {\ensuremath{\Lambda}}CDM}",
      journal = {\mnras},
         year = 2024,
        month = apr,
       volume = {529},
       number = {4},
        pages = {4611-4623},
          doi = {10.1093/mnras/stae833},
archivePrefix = {arXiv},
       eprint = {2308.16342},
 primaryClass = {astro-ph.CO},
       adsurl = {https://ui.adsabs.harvard.edu/abs/2024MNRAS.529.4611S}
}

@ARTICLE{Nadler_2025,
       author = {{Nadler}, Ethan O. and {Kong}, Demao and {Yang}, Daneng and {Yu}, Hai-Bo},
        title = "{SIDM Concerto: Compilation and Data Release of Self-interacting Dark Matter Zoom-in Simulations}",
      journal = {\apj},
         year = 2025,
        month = sep,
       volume = {991},
       number = {1},
          eid = {69},
        pages = {69},
          doi = {10.3847/1538-4357/adf553},
archivePrefix = {arXiv},
       eprint = {2503.10748},
 primaryClass = {astro-ph.CO},
       adsurl = {https://ui.adsabs.harvard.edu/abs/2025ApJ...991...69N}
}

@INPROCEEDINGS{Silverman_2025,
       author = {{Silverman}, Maya and {Kaplinghat}, Manoj and {Lisanti}, Mariangela and {Necib}, Lina and {Arora}, Arpit and {Hussein}, Abdelaziz},
        title = "{Mergers Matter: Gravothermal Collapse in Self-Interacting Dark Matter Halos}",
    booktitle = {246th Meeting of the AAS},
         year = 2025,
       volume = {246},
        month = jun,
          eid = {329.03D},
        pages = {329.03D},
       adsurl = {https://ui.adsabs.harvard.edu/abs/2025AAS...24632903S}
}

@ARTICLE{Silverman_2026,
       author = {{Silverman}, Maya and {Hussein}, Abdelaziz and {Arora}, Arpit and {Lisanti}, Mariangela and {Kaplinghat}, Manoj and {Necib}, Lina and {Thoyas}, Andreas and {O'Neil}, Stephanie and {Sanderson}, Robyn E. and {Shen}, Xuejian and {Moreno}, Jorge},
        title = "{Mergers Matter: Gravothermal Collapse in Dwarf Halos with Self-Interacting Dark Matter}",
      journal = {arXiv e-prints},
         year = 2026,
        month = jun,
          eid = {arXiv:2606.02566},
        pages = {arXiv:2606.02566},
          doi = {10.48550/arXiv.2606.02566},
archivePrefix = {arXiv},
       eprint = {2606.02566},
 primaryClass = {astro-ph.GA},
       adsurl = {https://ui.adsabs.harvard.edu/abs/2026arXiv260602566S}
}

@ARTICLE{Lei_2025,
       author = {{Lei}, Lei and {Wang}, Yi-Ying and {Li}, Qiao and {Dong}, Jiang and {Wang}, Ze-Fan and {Lin}, Wei-Long and {Shu}, Yi-Ping and {Cao}, Xiao-Yue and {Yang}, Da-Neng and {Fan}, Yi-Zhong},
        title = "{A Dense Dark Matter Core of the Subhalo in the Strong Lensing System JVAS B1938+666}",
      journal = {\apjl},
         year = 2025,
        month = sep,
       volume = {991},
       number = {1},
          eid = {L27},
        pages = {L27},
          doi = {10.3847/2041-8213/ae047c},
archivePrefix = {arXiv},
       eprint = {2509.07808},
 primaryClass = {astro-ph.CO},
       adsurl = {https://ui.adsabs.harvard.edu/abs/2025ApJ...991L..27L}
}

@ARTICLE{Harvey_2025,
       author = {{Harvey}, David and {Revaz}, Yves and {Schaller}, Matthieu and {Schneider}, Aurel and {Tregidga}, Ethan and {Vecchi}, Felix},
        title = "{DARKSKIES: A suite of super-sampled zoom-in simulations of galaxy clusters with self-interacting dark matter}",
      journal = {\aap},
         year = 2025,
        month = nov,
       volume = {703},
          eid = {A112},
        pages = {A112},
          doi = {10.1051/0004-6361/202555180},
archivePrefix = {arXiv},
       eprint = {2509.19842},
 primaryClass = {astro-ph.CO},
       adsurl = {https://ui.adsabs.harvard.edu/abs/2025A&A...703A.112H}
}

@ARTICLE{Straight_2025,
       author = {{Straight}, Maria C. and {Boylan-Kolchin}, Michael and {Bullock}, James S. and {Hopkins}, Philip F. and {Shen}, Xuejian and {Necib}, Lina and {Lazar}, Alexandres and {Graus}, Andrew S. and {Samuel}, Jenna},
        title = "{Central densities of dark matter haloes in FIRE-2 simulations of low-mass galaxies with cold dark matter and self-interacting dark matter}",
      journal = {\mnras},
         year = 2025,
        month = nov,
       volume = {543},
       number = {3},
        pages = {1995-2005},
          doi = {10.1093/mnras/staf1539},
archivePrefix = {arXiv},
       eprint = {2501.16602},
 primaryClass = {astro-ph.GA},
       adsurl = {https://ui.adsabs.harvard.edu/abs/2025MNRAS.543.1995S}
}

@ARTICLE{vanDenBosch2026,
       author = {{van den Bosch}, Frank C. and {Dattathri}, Shashank},
        title = "{Dynamics in the Cores of Self-Interacting Dark Matter Halos: Reduced Stalling and Accelerated Core Collapse}",
      journal = {OJAp},
         year = 2026,
        month = feb,
       volume = {9},
        pages = {57701},
          doi = {10.33232/001c.157701},
archivePrefix = {arXiv},
       eprint = {2511.14912},
 primaryClass = {astro-ph.GA},
       adsurl = {https://ui.adsabs.harvard.edu/abs/2026OJAp....957701V}
}

@ARTICLE{EP-Ragagnin,
       author = {{Euclid Collaboration: Ragagnin}, A. and {Saro}, A. and {Andreon}, S. and others},
        title = "{Euclid preparation: LXVI. Impact of line-of-sight projections on the covariance between galaxy cluster multi-wavelength observable properties: insights from hydrodynamic simulations}",
      journal = {\aap},
         year = 2025,
        month = mar,
       volume = {695},
          eid = {A282},
        pages = {A282},
          doi = {10.1051/0004-6361/202451347},
archivePrefix = {arXiv},
       eprint = {2412.00191},
 primaryClass = {astro-ph.CO},
       adsurl = {https://ui.adsabs.harvard.edu/abs/2025A&A...695A.282E}
}

@ARTICLE{Vogelsberger_2014,
       author = {{Vogelsberger}, Mark and {Zavala}, Jesus and {Simpson}, Christine and {Jenkins}, Adrian},
        title = "{Dwarf galaxies in CDM and SIDM with baryons: observational probes of the nature of dark matter}",
      journal = {\mnras},
         year = 2014,
        month = nov,
       volume = {444},
       number = {4},
        pages = {3684-3698},
          doi = {10.1093/mnras/stu1713},
archivePrefix = {arXiv},
       eprint = {1405.5216},
 primaryClass = {astro-ph.CO},
       adsurl = {https://ui.adsabs.harvard.edu/abs/2014MNRAS.444.3684V}
}

@ARTICLE{Yu_2026,
       author = {{Yu}, Hai-Bo},
        title = "{Core-Collapsed SIDM Halos as the Common Origin of Dense Perturbers in Lenses, Streams, and Satellites}",
      journal = {\prl},
         year = 2026,
        month = apr,
       volume = {136},
       number = {14},
          eid = {141001},
        pages = {141001},
          doi = {10.1103/txxx-97ln},
archivePrefix = {arXiv},
       eprint = {2510.11006},
 primaryClass = {astro-ph.GA},
       adsurl = {https://ui.adsabs.harvard.edu/abs/2026PhRvL.136n1001Y}
}

@ARTICLE{Despali_2022,
       author = {{Despali}, Giulia and {Walls}, Levi G. and {Vegetti}, Simona and {Sparre}, Martin and {Vogelsberger}, Mark and {Zavala}, Jes{\'u}s},
        title = "{Constraining SIDM with halo shapes: Revisited predictions from realistic simulations of early-type galaxies}",
      journal = {\mnras},
         year = 2022,
        month = nov,
       volume = {516},
       number = {3},
        pages = {4543-4559},
          doi = {10.1093/mnras/stac2521},
archivePrefix = {arXiv},
       eprint = {2204.12502},
 primaryClass = {astro-ph.CO},
       adsurl = {https://ui.adsabs.harvard.edu/abs/2022MNRAS.516.4543D}
}

@article{Planck_2020,
	author = {{Planck Collaboration} and {Aghanim, N.} and {Akrami, Y.} and {Arroja, F.} and {Ashdown, M.} and {Aumont, J.} and {Baccigalupi, C.} and {Ballardini, M.} and {Banday, A. J.} and {Barreiro, R. B.} and {Bartolo, N.} and {Basak, S.} and {Battye, R.} and {Benabed, K.} and {Bernard, J.-P.} and {Bersanelli, M.} and {Bielewicz, P.} and {Bock, J. J.} and {Bond, J. R.} and {Borrill, J.} and {Bouchet, F. R.} and {Boulanger, F.} and {Bucher, M.} and {Burigana, C.} and {Butler, R. C.} and {Calabrese, E.} and {Cardoso, J.-F.} and {Carron, J.} and {Casaponsa, B.} and {Challinor, A.} and {Chiang, H. C.} and {Colombo, L. P. L.} and {Combet, C.} and {Contreras, D.} and {Crill, B. P.} and {Cuttaia, F.} and {de Bernardis, P.} and {de Zotti, G.} and {Delabrouille, J.} and {Delouis, J.-M.} and {D\'esert, F.-X.} and {Di Valentino, E.} and {Dickinson, C.} and {Diego, J. M.} and {Donzelli, S.} and {Dor\'e, O.} and {Douspis, M.} and {Ducout, A.} and {Dupac, X.} and {Efstathiou, G.} and {Elsner, F.} and {En\ss{}lin, T. A.} and {Eriksen, H. K.} and {Falgarone, E.} and {Fantaye, Y.} and {Fergusson, J.} and {Fernandez-Cobos, R.} and {Finelli, F.} and {Forastieri, F.} and {Frailis, M.} and {Franceschi, E.} and {Frolov, A.} and {Galeotta, S.} and {Galli, S.} and {Ganga, K.} and {G\'enova-Santos, R. T.} and {Gerbino, M.} and {Ghosh, T.} and {Gonz\'alez-Nuevo, J.} and {G\'orski, K. M.} and {Gratton, S.} and {Gruppuso, A.} and {Gudmundsson, J. E.} and {Hamann, J.} and {Handley, W.} and {Hansen, F. K.} and {Helou, G.} and {Herranz, D.} and {Hildebrandt, S. R.} and {Hivon, E.} and {Huang, Z.} and {Jaffe, A. H.} and {Jones, W. C.} and {Karakci, A.} and {Keih\"anen, E.} and {Keskitalo, R.} and {Kiiveri, K.} and {Kim, J.} and {Kisner, T. S.} and {Knox, L.} and {Krachmalnicoff, N.} and {Kunz, M.} and {Kurki-Suonio, H.} and {Lagache, G.} and {Lamarre, J.-M.} and {Langer, M.} and {Lasenby, A.} and {Lattanzi, M.} and {Lawrence, C. R.} and {Le Jeune, M.} and {Leahy, J. P.} and {Lesgourgues, J.} and {Levrier, F.} and {Lewis, A.} and {Liguori, M.} and {Lilje, P. B.} and {Lilley, M.} and {Lindholm, V.} and {L\'opez-Caniego, M.} and {Lubin, P. M.} and {Ma, Y.-Z.} and {Mac\'{\i}as-P\'erez, J. F.} and {Maggio, G.} and {Maino, D.} and {Mandolesi, N.} and {Mangilli, A.} and {Marcos-Caballero, A.} and {Maris, M.} and {Martin, P. G.} and {Martinelli, M.} and {Mart\'{\i}nez-Gonz\'alez, E.} and {Matarrese, S.} and {Mauri, N.} and {McEwen, J. D.} and {Meerburg, P. D.} and {Meinhold, P. R.} and {Melchiorri, A.} and {Mennella, A.} and {Migliaccio, M.} and {Millea, M.} and {Mitra, S.} and {Miville-Desch\^enes, M.-A.} and {Molinari, D.} and {Moneti, A.} and {Montier, L.} and {Morgante, G.} and {Moss, A.} and {Mottet, S.} and {M\"unchmeyer, M.} and {Natoli, P.} and {N\o{}rgaard-Nielsen, H. U.} and {Oxborrow, C. A.} and {Pagano, L.} and {Paoletti, D.} and {Partridge, B.} and {Patanchon, G.} and {Pearson, T. J.} and {Peel, M.} and {Peiris, H. V.} and {Perrotta, F.} and {Pettorino, V.} and {Piacentini, F.} and {Polastri, L.} and {Polenta, G.} and {Puget, J.-L.} and {Rachen, J. P.} and {Reinecke, M.} and {Remazeilles, M.} and {Renault, C.} and {Renzi, A.} and {Rocha, G.} and {Rosset, C.} and {Roudier, G.} and {Rubi\~no-Mart\'{\i}n, J. A.} and {Ruiz-Granados, B.} and {Salvati, L.} and {Sandri, M.} and {Savelainen, M.} and {Scott, D.} and {Shellard, E. P. S.} and {Shiraishi, M.} and {Sirignano, C.} and {Sirri, G.} and {Spencer, L. D.} and {Sunyaev, R.} and {Suur-Uski, A.-S.} and {Tauber, J. A.} and {Tavagnacco, D.} and {Tenti, M.} and {Terenzi, L.} and {Toffolatti, L.} and {Tomasi, M.} and {Trombetti, T.} and {Valiviita, J.} and {Van Tent, B.} and {Vibert, L.} and {Vielva, P.} and {Villa, F.} and {Vittorio, N.} and {Wandelt, B. D.} and {Wehus, I. K.} and {White, M.} and {White, S. D. M.} and {Zacchei, A.} and {Zonca, A.}},
	title = {Planck 2018 results - I. Overview and the cosmological legacy of Planck},
	DOI= "10.1051/0004-6361/201833880",
	url= "https://doi.org/10.1051/0004-6361/201833880",
	journal = {\aap},
	year = 2020,
	volume = 641,
	pages = "A1",
    }

@article{Bullock_2017,
   author = "Bullock, James S. and Boylan-Kolchin, Michael",
   title = "Small-Scale Challenges to the ΛCDM Paradigm", 
   journal= "\araa",
   year = "2017",
   volume = "55",
   number = "Volume 55, 2017",
   pages = "343-387",
   doi = "https://doi.org/10.1146/annurev-astro-091916-055313",
   url = "https://www.annualreviews.org/content/journals/10.1146/annurev-astro-091916-055313",
   publisher = "Annual Reviews",
   issn = "1545-4282",
   type = "Journal Article",
  }

@article{Bozorgnia_2025,
   title={Dark matter candidates and searches},
   volume={103},
   ISSN={1208-6045},
   url={http://dx.doi.org/10.1139/cjp-2024-0128},
   DOI={10.1139/cjp-2024-0128},
   number={8},
   journal={Can. J. Phys.},
   publisher={Canadian Science Publishing},
   author={Bozorgnia, Nassim and Bramante, Joseph and Cline, James M. and Curtin, David and McKeen, David and Morrissey, David E. and Ritz, Adam and Viel, Simon and Vincent, Aaron C. and Zhang, Yue},
   year={2025},
   month=aug, pages={671–703} 
   }

@article{Spergel_2000,
  title = {Observational Evidence for Self-Interacting Cold Dark Matter},
  author = {Spergel, David N. and Steinhardt, Paul J.},
  journal = {Phys. Rev. Lett.},
  volume = {84},
  issue = {17},
  pages = {3760--3763},
  numpages = {0},
  year = {2000},
  month = {Apr},
  publisher = {American Physical Society},
  doi = {10.1103/PhysRevLett.84.3760},
  url = {https://link.aps.org/doi/10.1103/PhysRevLett.84.3760}
  }

@INPROCEEDINGS{Bertocco_2020,
       author = {{Bertocco}, S. and {Goz}, D. and {Tornatore}, L. and {Ragagnin}, A. and {Maggio}, G. and {Gasparo}, F. and {Vuerli}, C. and {Taffoni}, G. and {Molinaro}, M.},
        title = "{INAF Trieste Astronomical Observatory Information Technology Framework}",
    booktitle = {ASP},
         year = 2020,
       editor = {{Pizzo}, R. and {Deul}, E.~R. and {Mol}, J.~D. and {de Plaa}, J. and {Verkouter}, H.},
       series = {ASP Conference Series},
       volume = {527},
        month = jan,
        pages = {303},
archivePrefix = {arXiv},
       eprint = {1912.05340},
 primaryClass = {astro-ph.IM},
       adsurl = {https://ui.adsabs.harvard.edu/abs/2020ASPC..527..303B}
}

@INPROCEEDINGS{Taffoni_2020, 
       author = {{Taffoni}, G. and {Becciani}, U. and {Garilli}, B. and {Maggio}, G. and {Pasian}, F. and {Umana}, G. and {Smareglia}, R. and {Vitello}, F.},
        title = "{CHIPP: INAF Pilot Project for HTC, HPC and HPDA}",
    booktitle = {ASP},
         year = 2020,
       editor = {{Pizzo}, R. and {Deul}, E.~R. and {Mol}, J.~D. and {de Plaa}, J. and {Verkouter}, H.},
       series = {ASP Conference Series},
       volume = {527},
        month = jan,
        pages = {307},
archivePrefix = {arXiv},
       eprint = {2002.01283},
 primaryClass = {astro-ph.IM},
       adsurl = {https://ui.adsabs.harvard.edu/abs/2020ASPC..527..307T}
}

@Article{Harris_2020,
 title         = {Array programming with {NumPy}},
 author        = {Charles R. Harris and K. Jarrod Millman and St{\'{e}}fan J.
                 van der Walt and Ralf Gommers and Pauli Virtanen and David
                 Cournapeau and Eric Wieser and Julian Taylor and Sebastian
                 Berg and Nathaniel J. Smith and Robert Kern and Matti Picus
                 and Stephan Hoyer and Marten H. van Kerkwijk and Matthew
                 Brett and Allan Haldane and Jaime Fern{\'{a}}ndez del
                 R{\'{i}}o and Mark Wiebe and Pearu Peterson and Pierre
                 G{\'{e}}rard-Marchant and Kevin Sheppard and Tyler Reddy and
                 Warren Weckesser and Hameer Abbasi and Christoph Gohlke and
                 Travis E. Oliphant},
 year          = {2020},
 month         = sep,
 journal       = {Nat},
 volume        = {585},
 number        = {7825},
 pages         = {357--362},
 doi           = {10.1038/s41586-020-2649-2},
 publisher     = {Springer Science and Business Media {LLC}},
 url           = {https://doi.org/10.1038/s41586-020-2649-2}
}

@ARTICLE{Diemer_2018,
       author = {{Diemer}, Benedikt},
        title = "{COLOSSUS: A Python Toolkit for Cosmology, Large-scale Structure, and Dark Matter Halos}",
      journal = {\apjs},
         year = 2018,
        month = dec,
       volume = {239},
       number = {2},
          eid = {35},
        pages = {35},
          doi = {10.3847/1538-4365/aaee8c},
archivePrefix = {arXiv},
       eprint = {1712.04512},
 primaryClass = {astro-ph.CO},
       adsurl = {https://ui.adsabs.harvard.edu/abs/2018ApJS..239...35D}
}

@ARTICLE{Ludlow_2014,
       author = {{Ludlow}, Aaron D. and {Navarro}, Julio F. and {Angulo}, Ra{\'u}l E. and {Boylan-Kolchin}, Michael and {Springel}, Volker and {Frenk}, Carlos and {White}, Simon D.~M.},
        title = "{The mass-concentration-redshift relation of cold dark matter haloes}",
      journal = {\mnras},
         year = 2014,
        month = jun,
       volume = {441},
       number = {1},
        pages = {378-388},
          doi = {10.1093/mnras/stu483},
archivePrefix = {arXiv},
       eprint = {1312.0945},
 primaryClass = {astro-ph.CO},
       adsurl = {https://ui.adsabs.harvard.edu/abs/2014MNRAS.441..378L}
}

@article{Tulin_2018,
    title = {Dark matter self-interactions and small scale structure},
    journal = {Phys. Rep.},
    volume = {730},
    pages = {1-57},
    year = {2018},
    issn = {0370-1573},
    doi = {https://doi.org/10.1016/j.physrep.2017.11.004},
    url = {https://www.sciencedirect.com/science/article/pii/S0370157317304039},
    author = {Sean Tulin and Hai-Bo Yu}
    }

@ARTICLE{Eckert_2022,
       author = {{Eckert}, D. and {Ettori}, S. and {Robertson}, A. and {Massey}, R. and {Pointecouteau}, E. and {Harvey}, D. and {McCarthy}, I.~G.},
        title = "{Constraints on dark matter self-interaction from the internal density profiles of X-COP galaxy clusters}",
      journal = {\aap},
         year = 2022,
        month = oct,
       volume = {666},
          eid = {A41},
        pages = {A41},
          doi = {10.1051/0004-6361/202243205},
archivePrefix = {arXiv},
       eprint = {2205.01123},
 primaryClass = {astro-ph.CO},
       adsurl = {https://ui.adsabs.harvard.edu/abs/2022A&A...666A..41E}
}

@article{DESI_Collaboration_2022,
   title={Overview of the Instrumentation for the Dark Energy Spectroscopic Instrument},
   volume={164},
   ISSN={1538-3881},
   url={http://dx.doi.org/10.3847/1538-3881/ac882b},
   DOI={10.3847/1538-3881/ac882b},
   number={5},
   journal={\aj},
   publisher={American Astronomical Society},
   author={{DESI Collaboration} and Abareshi, B. and Aguilar, J. and Ahlen, S. and Alam, Shadab and Alexander, David M. and Alfarsy, R. and Allen, L. and Allende Prieto, C. and Alves, O. and Ameel, J. and Armengaud, E. and Asorey, J. and Aviles, Alejandro and Bailey, S. and Balaguera-Antolínez, A. and Ballester, O. and Baltay, C. and Bault, A. and Beltran, S. F. and Benavides, B. and BenZvi, S. and Berti, A. and Besuner, R. and Beutler, Florian and Bianchi, D. and Blake, C. and Blanc, P. and Blum, R. and Bolton, A. and Bose, S. and Bramall, D. and Brieden, S. and Brodzeller, A. and Brooks, D. and Brownewell, C. and Buckley-Geer, E. and Cahn, R. N. and Cai, Z. and Canning, R. and Capasso, R. and Carnero Rosell, A. and Carton, P. and Casas, R. and Castander, F. J. and Cervantes-Cota, J. L. and Chabanier, S. and Chaussidon, E. and Chuang, C. and Circosta, C. and Cole, S. and Cooper, A. P. and da Costa, L. and Cousinou, M.-C. and Cuceu, A. and Davis, T. M. and Dawson, K. and de la Cruz-Noriega, R. and de la Macorra, A. and de Mattia, A. and Della Costa, J. and Demmer, P. and Derwent, M. and Dey, A. and Dey, B. and Dhungana, G. and Ding, Z. and Dobson, C. and Doel, P. and Donald-McCann, J. and Donaldson, J. and Douglass, K. and Duan, Y. and Dunlop, P. and Edelstein, J. and Eftekharzadeh, S. and Eisenstein, D. J. and Enriquez-Vargas, M. and Escoffier, S. and Evatt, M. and Fagrelius, P. and Fan, X. and Fanning, K. and Fawcett, V. A. and Ferraro, S. and Ereza, J. and Flaugher, B. and Font-Ribera, A. and Forero-Romero, J. E. and Frenk, C. S. and Fromenteau, S. and Gänsicke, B. T. and Garcia-Quintero, C. and Garrison, L. and Gaztañaga, E. and Gerardi, F. and Gil-Marín, H. and Gontcho A Gontcho, S. and Gonzalez-Morales, Alma X. and Gonzalez-de-Rivera, G. and Gonzalez-Perez, V. and Gordon, C. and Graur, O. and Green, D. and Grove, C. and Gruen, D. and Gutierrez, G. and Guy, J. and Hahn, C. and Harris, S. and Herrera, D. and Herrera-Alcantar, Hiram K. and Honscheid, K. and Howlett, C. and Huterer, D. and Iršič, V. and Ishak, M. and Jelinsky, P. and Jiang, L. and Jimenez, J. and Jing, Y. P. and Joyce, R. and Jullo, E. and Juneau, S. and Karaçaylı, N. G. and Karamanis, M. and Karcher, A. and Karim, T. and Kehoe, R. and Kent, S. and Kirkby, D. and Kisner, T. and Kitaura, F. and Koposov, S. E. and Kovács, A. and Kremin, A. and Krolewski, Alex and L’Huillier, B. and Lahav, O. and Lambert, A. and Lamman, C. and Lan, Ting-Wen and Landriau, M. and Lane, S. and Lang, D. and Lange, J. U. and Lasker, J. and Le Guillou, L. and Leauthaud, A. and Le Van Suu, A. and Levi, Michael E. and Li, T. S. and Magneville, C. and Manera, M. and Manser, Christopher J. and Marshall, B. and Martini, Paul and McCollam, W. and McDonald, P. and Meisner, Aaron M. and Mena-Fernández, J. and Meneses-Rizo, J. and Mezcua, M. and Miller, T. and Miquel, R. and Montero-Camacho, P. and Moon, J. and Moustakas, J. and Mueller, E. and Muñoz-Gutiérrez, Andrea and Myers, Adam D. and Nadathur, S. and Najita, J. and Napolitano, L. and Neilsen, E. and Newman, Jeffrey A. and Nie, J. D. and Ning, Y. and Niz, G. and Norberg, P. and Noriega, Hernán E. and O’Brien, T. and Obuljen, A. and Palanque-Delabrouille, N. and Palmese, A. and Zhiwei, P. and Pappalardo, D. and PENG, X. and Percival, W. J. and Perruchot, S. and Pogge, R. and Poppett, C. and Porredon, A. and Prada, F. and Prochaska, J. and Pucha, R. and Pérez-Fernández, A. and Pérez-Ràfols, I. and Rabinowitz, D. and Raichoor, A. and Ramirez-Solano, S. and Ramírez-Pérez, César and Ravoux, C. and Reil, K. and Rezaie, M. and Rocher, A. and Rockosi, C. and Roe, N. A. and Roodman, A. and Ross, A. J. and Rossi, G. and Ruggeri, R. and Ruhlmann-Kleider, V. and Sabiu, C. G. and Gaines, S. and Said, K. and Saintonge, A. and Salas Catonga, Javier and Samushia, L. and Sanchez, E. and Saulder, C. and Schaan, E. and Schlafly, E. and Schlegel, D. and Schmoll, J. and Scholte, D. and Schubnell, M. and Secroun, A. and Seo, H. and Serrano, S. and Sharples, Ray M. and Sholl, Michael J. and Silber, Joseph Harry and Silva, D. R. and Sirk, M. and Siudek, M. and Smith, A. and Sprayberry, D. and Staten, R. and Stupak, B. and Tan, T. and Tarlé, Gregory and Tie, Suk Sien and Tojeiro, R. and Ureña-López, L. A. and Valdes, F. and Valenzuela, O. and Valluri, M. and Vargas-Magaña, M. and Verde, L. and Walther, M. and Wang, B. and Wang, M. S. and Weaver, B. A. and Weaverdyck, C. and Wechsler, R. and Wilson, Michael J. and Yang, J. and Yu, Y. and Yuan, S. and Yèche, Christophe and Zhang, H. and Zhang, K. and Zhao, Cheng and Zhou, Rongpu and Zhou, Zhimin and Zou, H. and Zou, J. and Zou, S. and Zu, Y.},
   year={2022},
   month=oct, 
   pages={207} 
   }

@article{Navarro_1997,
   title={A Universal Density Profile from Hierarchical Clustering},
   volume={490},
   ISSN={1538-4357},
   url={http://dx.doi.org/10.1086/304888},
   DOI={10.1086/304888},
   number={2},
   journal={\apj},
   publisher={American Astronomical Society},
   author={Navarro, Julio F. and Frenk, Carlos S. and White, Simon D. M.},
   year={1997},
   month=dec, 
   pages={493–508} 
   }

@article{Oman_2015,
    author = {Oman, Kyle A. and Navarro, Julio F. and Fattahi, Azadeh and Frenk, Carlos S. and Sawala, Till and White, Simon D. M. and Bower, Richard and Crain, Robert A. and Furlong, Michelle and Schaller, Matthieu and Schaye, Joop and Theuns, Tom},
    title = {The unexpected diversity of dwarf galaxy rotation curves},
    journal = {\mnras},
    volume = {452},
    number = {4},
    pages = {3650-3665},
    year = {2015},
    month = {10},
    issn = {0035-8711},
    doi = {10.1093/mnras/stv1504},
    url = {https://doi.org/10.1093/mnras/stv1504},
    eprint = {https://academic.oup.com/mnras/article-pdf/452/4/3650/18237239/stv1504.pdf},
    }

@article{de_Blok_2009,
   title={The Core‐Cusp Problem},
   volume={2010},
   ISSN={1687-7977},
   url={http://dx.doi.org/10.1155/2010/789293},
   DOI={10.1155/2010/789293},
   number={1},
   journal={Adv. Astron.},
   publisher={Wiley},
   author={de Blok, W. J. G.},
   editor={Brinks, Elias},
   year={2009},
   month=nov
   }

@ARTICLE{EuclidSkyOverview,
author = {{Euclid Collaboration: Mellier}, Y. and {Abdurro'uf} and {Acevedo~Barroso}, J.A. and others},
	title = {Euclid - I. Overview of the Euclid mission},
	DOI= "10.1051/0004-6361/202450810",
	url= "https://doi.org/10.1051/0004-6361/202450810",
	journal = {\aap},
	year = 2025,
	volume = 697,
	pages = "A1",
}

@ARTICLE{Dolag_2005,
       author = {{Dolag}, K. and {Vazza}, F. and {Brunetti}, G. and {Tormen}, G.},
        title = "{Turbulent gas motions in galaxy cluster simulations: the role of smoothed particle hydrodynamics viscosity}",
      journal = {\mnras},
         year = 2005,
        month = dec,
       volume = {364},
       number = {3},
        pages = {753-772},
          doi = {10.1111/j.1365-2966.2005.09630.x},
archivePrefix = {arXiv},
       eprint = {astro-ph/0507480},
 primaryClass = {astro-ph},
       adsurl = {https://ui.adsabs.harvard.edu/abs/2005MNRAS.364..753D}
}

@article{Dehnen_Aly_2012,
    author = {Dehnen, Walter and Aly, Hossam},
    title = {Improving convergence in smoothed particle hydrodynamics simulations without pairing instability},
    journal = {\mnras},
    volume = {425},
    number = {2},
    pages = {1068-1082},
    year = {2012},
    month = {09},
    issn = {0035-8711},
    doi = {10.1111/j.1365-2966.2012.21439.x},
    url = {https://doi.org/10.1111/j.1365-2966.2012.21439.x},
    eprint = {https://academic.oup.com/mnras/article-pdf/425/2/1068/4013442/425-2-1068.pdf},
}

@ARTICLE{Dolag_2004,
       author = {{Dolag}, K. and {Jubelgas}, M. and {Springel}, V. and {Borgani}, S. and {Rasia}, E.},
        title = "{Thermal Conduction in Simulated Galaxy Clusters}",
      journal = {\apjl},
         year = 2004,
        month = may,
       volume = {606},
       number = {2},
        pages = {L97-L100},
          doi = {10.1086/420966},
archivePrefix = {arXiv},
       eprint = {astro-ph/0401470},
 primaryClass = {astro-ph},
       adsurl = {https://ui.adsabs.harvard.edu/abs/2004ApJ...606L..97D}
}

@ARTICLE{Springel_DiMatteo_2005,
       author = {{Springel}, Volker and {Di Matteo}, Tiziana and {Hernquist}, Lars},
        title = "{Black Holes in Galaxy Mergers: The Formation of Red Elliptical Galaxies}",
      journal = {\apjl},
         year = 2005,
        month = feb,
       volume = {620},
       number = {2},
        pages = {L79-L82},
          doi = {10.1086/428772},
archivePrefix = {arXiv},
       eprint = {astro-ph/0409436},
 primaryClass = {astro-ph},
       adsurl = {https://ui.adsabs.harvard.edu/abs/2005ApJ...620L..79S}
}

@article{Ragagnin_2024,
	author = {{Ragagnin, A.} and {Meneghetti, M.} and {Calura, F.} and {Despali, G.} and {Dolag, K.} and {Fischer, M. S.} and {Giocoli, C.} and {Moscardini, L.}},
	title = {Dianoga SIDM: Galaxy cluster self-interacting dark matter simulations},
	DOI= "10.1051/0004-6361/202449872",
	url= "https://doi.org/10.1051/0004-6361/202449872",
	journal = {\aap},
	year = 2024,
	volume = 687,
	pages = "A270",
}

@article{Postman_2012,
   title={THE CLUSTER LENSING AND SUPERNOVA SURVEY WITH HUBBLE: AN OVERVIEW},
   volume={199},
   ISSN={1538-4365},
   url={http://dx.doi.org/10.1088/0067-0049/199/2/25},
   DOI={10.1088/0067-0049/199/2/25},
   number={2},
   journal={\apjs},
   publisher={American Astronomical Society},
   author={Postman, Marc and Coe, Dan and Benítez, Narciso and Bradley, Larry and Broadhurst, Tom and Donahue, Megan and Ford, Holland and Graur, Or and Graves, Genevieve and Jouvel, Stephanie and Koekemoer, Anton and Lemze, Doron and Medezinski, Elinor and Molino, Alberto and Moustakas, Leonidas and Ogaz, Sara and Riess, Adam and Rodney, Steve and Rosati, Piero and Umetsu, Keiichi and Zheng, Wei and Zitrin, Adi and Bartelmann, Matthias and Bouwens, Rychard and Czakon, Nicole and Golwala, Sunil and Host, Ole and Infante, Leopoldo and Jha, Saurabh and Jimenez-Teja, Yolanda and Kelson, Daniel and Lahav, Ofer and Lazkoz, Ruth and Maoz, Dani and McCully, Curtis and Melchior, Peter and Meneghetti, Massimo and Merten, Julian and Moustakas, John and Nonino, Mario and Patel, Brandon and Regös, Enikö and Sayers, Jack and Seitz, Stella and Van der Wel, Arjen},
   year={2012},
   month=mar, pages={25} 
   }

@article{Adhikari_2025,
   title={Astrophysical tests of dark matter self-interactions},
   volume={97},
   ISSN={1539-0756},
   url={http://dx.doi.org/10.1103/m2vm-59y3},
   DOI={10.1103/m2vm-59y3},
   number={4},
   journal={Rev. Mod. Phys.},
   publisher={American Physical Society (APS)},
   author={Adhikari, Susmita and Banerjee, Arka and Boddy, Kimberly K. and Cyr-Racine, Francis-Yan and Desmond, Harry and Dvorkin, Cora and Jain, Bhuvnesh and Kahlhoefer, Felix and Kaplinghat, Manoj and Nierenberg, Anna and Peter, Annika H. G. and Robertson, Andrew and Sakstein, Jeremy and Zavala, Jesús},
   year={2025},
   month=dec 
   }

@ARTICLE{EP-Racz,
       author = {{Euclid Collaboration: Racz}, G. and {Breton}, M.-A. and {Fiorini}, B. and {Le Brun}, A.~M.~C. and {Winther}, H.-A. and {Sakr}, Z. and {Pizzuti}, L. and {Ragagnin}, A. and {Gayoux}, T. and {Altamura}, E. and {Carella}, E. and {Pardede}, K. and {Verza}, G. and {Koyama}, K.},
        title = "{Euclid preparation: LXIII. Simulations and non-linearities beyond Lambda cold dark matter. 2. Results from non-standard simulations}",
      journal = {\aap},
         year = 2025,
        month = mar,
       volume = {695},
          eid = {A232},
        pages = {A232},
          doi = {10.1051/0004-6361/202452185},
archivePrefix = {arXiv},
       eprint = {2409.03523},
 primaryClass = {astro-ph.CO},
       adsurl = {https://ui.adsabs.harvard.edu/abs/2025A&A...695A.232E}
}

@ARTICLE{Fischer_2023,
       author = {{Fischer}, Moritz S. and {Durke}, Nils-Henrik and {Hollingshausen}, Katharina and {Hammer}, Claudius and {Br{\"u}ggen}, Marcus and {Dolag}, Klaus},
        title = "{The role of baryons in self-interacting dark matter mergers}",
      journal = {\mnras},
         year = 2023,
        month = aug,
       volume = {523},
       number = {4},
        pages = {5915-5933},
          doi = {10.1093/mnras/stad1786},
archivePrefix = {arXiv},
       eprint = {2302.07882},
 primaryClass = {astro-ph.CO},
       adsurl = {https://ui.adsabs.harvard.edu/abs/2023MNRAS.523.5915F}
}

@article{Fischer_2024a,
   title={Cosmological and idealized simulations of dark matter haloes with velocity-dependent, rare and frequent self-interactions},
   volume={529},
   ISSN={1365-2966},
   url={http://dx.doi.org/10.1093/mnras/stae699},
   DOI={10.1093/mnras/stae699},
   number={3},
   journal={\mnras},
   publisher={Oxford University Press (OUP)},
   author={Fischer, Moritz S and Kasselmann, Lenard and Brüggen, Marcus and Dolag, Klaus and Kahlhoefer, Felix and Ragagnin, Antonio and Robertson, Andrew and Schmidt-Hoberg, Kai},
   year={2024},
   month=mar, 
   pages={2327–2348} 
   }

@ARTICLE{Fischer_2024b,
       author = {{Fischer}, Moritz S. and {Dolag}, Klaus and {Yu}, Hai-Bo},
        title = "{Numerical challenges for energy conservation in N-body simulations of collapsing self-interacting dark matter halos}",
      journal = {\aap},
         year = 2024,
        month = sep,
       volume = {689},
          eid = {A300},
        pages = {A300},
          doi = {10.1051/0004-6361/202449849},
archivePrefix = {arXiv},
       eprint = {2403.00739},
 primaryClass = {astro-ph.CO},
       adsurl = {https://ui.adsabs.harvard.edu/abs/2024A&A...689A.300F}
}

@ARTICLE{Fischer_2025,
       author = {{Fischer}, Moritz S. and {Yu}, Hai-Bo and {Dolag}, Klaus},
        title = "{Accurately simulating core-collapse self-interacting dark matter halos}",
      journal = {\aap},
         year = 2025,
        month = nov,
       volume = {703},
          eid = {A234},
        pages = {A234},
          doi = {10.1051/0004-6361/202556189},
archivePrefix = {arXiv},
       eprint = {2506.06269},
 primaryClass = {astro-ph.CO},
       adsurl = {https://ui.adsabs.harvard.edu/abs/2025A&A...703A.234F}
}

@article{Garrison-Kimmel_2013,
    author = "Garrison-Kimmel, Shea and Rocha, Miguel and Boylan-Kolchin, Michael and Bullock, James and Lally, Jaspreet",
    title = "{Can Feedback Solve the Too Big to Fail Problem?}",
    eprint = "1301.3137",
    archivePrefix = "arXiv",
    primaryClass = "astro-ph.CO",
    doi = "10.1093/mnras/stt984",
    journal = {\mnras},
    volume = "433",
    pages = "3539",
    year = "2013"
}

@article{Loeb_2011,
   title={Cores in Dwarf Galaxies from Dark Matter with a Yukawa Potential},
   volume={106},
   ISSN={1079-7114},
   url={http://dx.doi.org/10.1103/PhysRevLett.106.171302},
   DOI={10.1103/physrevlett.106.171302},
   number={17},
   journal={Phys. Rev. Lett.},
   publisher={American Physical Society (APS)},
   author={Loeb, Abraham and Weiner, Neal},
   year={2011},
   month=apr 
   }

@article{Silverman_2023,
    author = {Silverman, Maya and Bullock, James S and Kaplinghat, Manoj and Robles, Victor H and Valli, Mauro},
    title = {Motivations for a large self-interacting dark matter cross-section from Milky Way satellites},
    journal = {\mnras},
    volume = {518},
    number = {2},
    pages = {2418-2435},
    year = {2023},
    month = {01},
    issn = {0035-8711},
    doi = {10.1093/mnras/stac3232},
    url = {https://doi.org/10.1093/mnras/stac3232},
    eprint = {https://academic.oup.com/mnras/article-pdf/518/2/2418/47266606/stac3232.pdf},
    }

@article{Lovell_2023,
    author = {Lovell, Mark R and Zavala, Jesús},
    title = {Matching the mass function of Milky Way satellites in competing dark matter models},
    journal = {\mnras},
    volume = {520},
    number = {1},
    pages = {1567-1589},
    year = {2023},
    month = {03},
    issn = {0035-8711},
    doi = {10.1093/mnras/stad216},
    url = {https://doi.org/10.1093/mnras/stad216},
    eprint = {https://academic.oup.com/mnras/article-pdf/520/1/1567/49116750/stad216.pdf},
    }

@ARTICLE{Schmidt_2026,
       author = {{Schmidt}, Ludwig D. and {Fischer}, Moritz S. and {Garny}, Mathias},
        title = "{Cooling, conduction, compact objects: Gravothermal evolution of dissipative self-interacting dark matter halos}",
      journal = {arXiv e-prints},
         year = 2026,
        month = jun,
          eid = {arXiv:2606.19428},
        pages = {arXiv:2606.19428},
          doi = {10.48550/arXiv.2606.19428},
archivePrefix = {arXiv},
       eprint = {2606.19428},
 primaryClass = {astro-ph.CO},
       adsurl = {https://ui.adsabs.harvard.edu/abs/2026arXiv260619428S}
}

@ARTICLE{Springel_2005,
       author = {{Springel}, Volker},
        title = "{The cosmological simulation code GADGET-2}",
      journal = {\mnras},
         year = 2005,
        month = dec,
       volume = {364},
       number = {4},
        pages = {1105-1134},
          doi = {10.1111/j.1365-2966.2005.09655.x},
archivePrefix = {arXiv},
       eprint = {astro-ph/0505010},
 primaryClass = {astro-ph},
       adsurl = {https://ui.adsabs.harvard.edu/abs/2005MNRAS.364.1105S}
    }

@article{Beck_2016,
    author = {Beck, A. M. and Murante, G. and Arth, A. and Remus, R.-S. and Teklu, A. F. and Donnert, J. M. F. and Planelles, S. and Beck, M. C. and Förster, P. and Imgrund, M. and Dolag, K. and Borgani, S.},
    title = {An improved SPH scheme for cosmological simulations},
    journal = {\mnras},
    volume = {455},
    number = {2},
    pages = {2110-2130},
    year = {2016},
    month = {01},
    issn = {0035-8711},
    doi = {10.1093/mnras/stv2443},
    url = {https://doi.org/10.1093/mnras/stv2443},
    eprint = {https://academic.oup.com/mnras/article-pdf/455/2/2110/18514729/stv2443.pdf},
    }

@article{Tornatore_2007,
    author = {Tornatore, L. and Borgani, S. and Dolag, K. and Matteucci, F.},
    title = {Chemical enrichment of galaxy clusters from hydrodynamical simulations},
    journal = {\mnras},
    volume = {382},
    number = {3},
    pages = {1050-1072},
    year = {2007},
    month = {11},
    issn = {0035-8711},
    doi = {10.1111/j.1365-2966.2007.12070.x},
    url = {https://doi.org/10.1111/j.1365-2966.2007.12070.x},
    eprint = {https://academic.oup.com/mnras/article-pdf/382/3/1050/18422138/mnras0382-1050.pdf},
    }

@ARTICLE{Davis_1985,
       author = {{Davis}, M. and {Efstathiou}, G. and {Frenk}, C.~S. and {White}, S.~D.~M.},
        title = "{The evolution of large-scale structure in a universe dominated by cold dark matter}",
      journal = {\apj},
         year = 1985,
        month = may,
       volume = {292},
        pages = {371-394},
          doi = {10.1086/163168},
       adsurl = {https://ui.adsabs.harvard.edu/abs/1985ApJ...292..371D}
    }

@article{Dolag_2009,
    author = {Dolag, K. and Borgani, S. and Murante, G. and Springel, V.},
    title = {Substructures in hydrodynamical cluster simulations},
    journal = {\mnras},
    volume = {399},
    number = {2},
    pages = {497-514},
    year = {2009},
    month = {10},
    issn = {0035-8711},
    doi = {10.1111/j.1365-2966.2009.15034.x},
    url = {https://doi.org/10.1111/j.1365-2966.2009.15034.x},
    eprint = {https://academic.oup.com/mnras/article-pdf/399/2/497/3620682/mnras0399-0497.pdf},
    }

@ARTICLE{Girardi2015,
       author = {{Girardi}, M. and {Mercurio}, A. and {Balestra}, I. and {Nonino}, M. and {Biviano}, A. and {Grillo}, C. and {Rosati}, P. and {Annunziatella}, M. and {Demarco}, R. and {Fritz}, A. and {Gobat}, R. and {Lemze}, D. and {Presotto}, V. and {Scodeggio}, M. and {Tozzi}, P. and {Bartosch Caminha}, G. and {Brescia}, M. and {Coe}, D. and {Kelson}, D. and {Koekemoer}, A. and {Lombardi}, M. and {Medezinski}, E. and {Postman}, M. and {Sartoris}, B. and {Umetsu}, K. and {Zitrin}, A. and {Boschin}, W. and {Czoske}, O. and {De Lucia}, G. and {Kuchner}, U. and {Maier}, C. and {Meneghetti}, M. and {Monaco}, P. and {Monna}, A. and {Munari}, E. and {Seitz}, S. and {Verdugo}, M. and {Ziegler}, B.},
        title = "{CLASH-VLT: Substructure in the galaxy cluster MACS J1206.2-0847 from kinematics of galaxy populations}",
      journal = {\aap},
         year = 2015,
        month = jul,
       volume = {579},
          eid = {A4},
        pages = {A4},
          doi = {10.1051/0004-6361/201425599},
archivePrefix = {arXiv},
       eprint = {1503.05607},
 primaryClass = {astro-ph.CO},
       adsurl = {https://ui.adsabs.harvard.edu/abs/2015A&A...579A...4G}
}

@ARTICLE{Pizzuti_2022,
       author = {{Pizzuti}, Lorenzo and {Saltas}, Ippocratis D. and {Umetsu}, Keiichi and {Sartoris}, Barbara},
        title = "{Probing vainsthein-screening gravity with galaxy clusters using internal kinematics and strong and weak lensing}",
      journal = {\mnras},
         year = 2022,
        month = may,
       volume = {512},
       number = {3},
        pages = {4280-4290},
          doi = {10.1093/mnras/stac746},
archivePrefix = {arXiv},
       eprint = {2112.12139},
 primaryClass = {astro-ph.CO},
       adsurl = {https://ui.adsabs.harvard.edu/abs/2022MNRAS.512.4280P}
}

@ARTICLE{RagoneFigueroa_2018,
       author = {{Ragone-Figueroa}, C. and {Granato}, G.~L. and {Ferraro}, M.~E. and {Murante}, G. and {Biffi}, V. and {Borgani}, S. and {Planelles}, S. and {Rasia}, E.},
        title = "{BCG mass evolution in cosmological hydro-simulations}",
      journal = {\mnras},
         year = 2018,
        month = sep,
       volume = {479},
       number = {1},
        pages = {1125-1136},
          doi = {10.1093/mnras/sty1639},
archivePrefix = {arXiv},
       eprint = {1803.08049},
 primaryClass = {astro-ph.GA},
       adsurl = {https://ui.adsabs.harvard.edu/abs/2018MNRAS.479.1125R}
}

@ARTICLE{Ferland_1998,
       author = {{Ferland}, G.~J. and {Korista}, K.~T. and {Verner}, D.~A. and {Ferguson}, J.~W. and {Kingdon}, J.~B. and {Verner}, E.~M.},
        title = "{CLOUDY 90: Numerical Simulation of Plasmas and Their Spectra}",
      journal = {\pasp},
         year = 1998,
        month = jul,
       volume = {110},
       number = {749},
        pages = {761-778},
          doi = {10.1086/316190},
       adsurl = {https://ui.adsabs.harvard.edu/abs/1998PASP..110..761F}
}

@ARTICLE{Price_2012,
       author = {{Price}, Daniel J.},
        title = "{Smoothed particle hydrodynamics and magnetohydrodynamics}",
      journal = {J. Comput. Phys.},
         year = 2012,
        month = feb,
       volume = {231},
       number = {3},
        pages = {759-794},
          doi = {10.1016/j.jcp.2010.12.011},
archivePrefix = {arXiv},
       eprint = {1012.1885},
 primaryClass = {astro-ph.IM},
       adsurl = {https://ui.adsabs.harvard.edu/abs/2012JCoPh.231..759P}
}

@ARTICLE{Hirschmann_2014,
       author = {{Hirschmann}, Michaela and {Dolag}, Klaus and {Saro}, Alexandro and {Bachmann}, Lisa and {Borgani}, Stefano and {Burkert}, Andreas},
        title = "{Cosmological simulations of black hole growth: AGN luminosities and downsizing}",
      journal = {\mnras},
         year = 2014,
        month = aug,
       volume = {442},
       number = {3},
        pages = {2304-2324},
          doi = {10.1093/mnras/stu1023},
archivePrefix = {arXiv},
       eprint = {1308.0333},
 primaryClass = {astro-ph.CO},
       adsurl = {https://ui.adsabs.harvard.edu/abs/2014MNRAS.442.2304H}
}

@ARTICLE{Fabjan_2010,
       author = {{Fabjan}, D. and {Borgani}, S. and {Tornatore}, L. and {Saro}, A. and {Murante}, G. and {Dolag}, K.},
        title = "{Simulating the effect of active galactic nuclei feedback on the metal enrichment of galaxy clusters}",
      journal = {\mnras},
         year = 2010,
        month = jan,
       volume = {401},
       number = {3},
        pages = {1670-1690},
          doi = {10.1111/j.1365-2966.2009.15794.x},
archivePrefix = {arXiv},
       eprint = {0909.0664},
 primaryClass = {astro-ph.CO},
       adsurl = {https://ui.adsabs.harvard.edu/abs/2010MNRAS.401.1670F}
}

@article{Dolag_2025,
       author = {{Dolag}, Klaus and {Remus}, Rhea-Silvia and {Valenzuela}, Lucas M. and {Kimmig}, Lucas C. and {Seidel}, Benjamin and {Fortune}, Silvio and {Stoiber}, Johannes and {Ivleva}, Anna and {Hoffmann}, Tadziu and {Biffi}, Veronica and {Marini}, Ilaria and {Popesso}, Paola and {Vladutescu-Zopp}, Stephan},
        title = "{Encyclopedia Magneticum: Scaling Relations from Cosmic Dawn to Present Day}",
      journal = {\aap},
         year = 2026,
         pages = {in press},
          eid = {arXiv:2504.01061},
        pages = {arXiv:2504.01061},
          doi = {10.48550/arXiv.2504.01061},
archivePrefix = {arXiv},
       eprint = {2504.01061},
 primaryClass = {astro-ph.CO},
       adsurl = {https://ui.adsabs.harvard.edu/abs/2025arXiv250401061D}
}

@ARTICLE{Tornatore_2010,
       author = {{Tornatore}, L. and {Borgani}, S. and {Viel}, M. and {Springel}, V.},
        title = "{The impact of feedback on the low-redshift intergalactic medium}",
      journal = {\mnras},
         year = 2010,
        month = mar,
       volume = {402},
       number = {3},
        pages = {1911-1926},
          doi = {10.1111/j.1365-2966.2009.16025.x},
archivePrefix = {arXiv},
       eprint = {0911.0699},
 primaryClass = {astro-ph.GA},
       adsurl = {https://ui.adsabs.harvard.edu/abs/2010MNRAS.402.1911T}
}

@ARTICLE{Wiersma_2009,
       author = {{Wiersma}, Robert P.~C. and {Schaye}, Joop and {Theuns}, Tom and {Dalla Vecchia}, Claudio and {Tornatore}, Luca},
        title = "{Chemical enrichment in cosmological, smoothed particle hydrodynamics simulations}",
      journal = {\mnras},
         year = 2009,
        month = oct,
       volume = {399},
       number = {2},
        pages = {574-600},
          doi = {10.1111/j.1365-2966.2009.15331.x},
archivePrefix = {arXiv},
       eprint = {0902.1535},
 primaryClass = {astro-ph.CO},
       adsurl = {https://ui.adsabs.harvard.edu/abs/2009MNRAS.399..574W}
}

@ARTICLE{Springel_Hernquist_2003,
       author = {{Springel}, Volker and {Hernquist}, Lars},
        title = "{Cosmological smoothed particle hydrodynamics simulations: a hybrid multiphase model for star formation}",
      journal = {\mnras},
         year = 2003,
        month = feb,
       volume = {339},
       number = {2},
        pages = {289-311},
          doi = {10.1046/j.1365-8711.2003.06206.x},
archivePrefix = {arXiv},
       eprint = {astro-ph/0206393},
 primaryClass = {astro-ph},
       adsurl = {https://ui.adsabs.harvard.edu/abs/2003MNRAS.339..289S}
}

@ARTICLE{Tornatore_2004,
       author = {{Tornatore}, L. and {Borgani}, S. and {Matteucci}, F. and {Recchi}, S. and {Tozzi}, P.},
        title = "{Simulating the metal enrichment of the intracluster medium}",
      journal = {\mnras},
         year = 2004,
        month = mar,
       volume = {349},
       number = {1},
        pages = {L19-L24},
          doi = {10.1111/j.1365-2966.2004.07689.x},
archivePrefix = {arXiv},
       eprint = {astro-ph/0401576},
 primaryClass = {astro-ph},
       adsurl = {https://ui.adsabs.harvard.edu/abs/2004MNRAS.349L..19T}
}

@ARTICLE{Rasia_2015,
       author = {{Rasia}, E. and {Borgani}, S. and {Murante}, G. and {Planelles}, S. and {Beck}, A.~M. and {Biffi}, V. and {Ragone-Figueroa}, C. and {Granato}, G.~L. and {Steinborn}, L.~K. and {Dolag}, K.},
        title = "{Cool Core Clusters from Cosmological Simulations}",
      journal = {\apjl},
         year = 2015,
        month = nov,
       volume = {813},
       number = {1},
          eid = {L17},
        pages = {L17},
          doi = {10.1088/2041-8205/813/1/L17},
archivePrefix = {arXiv},
       eprint = {1509.04247},
 primaryClass = {astro-ph.CO},
       adsurl = {https://ui.adsabs.harvard.edu/abs/2015ApJ...813L..17R}
}

@article{Chaitra_2026,
    title = {Variability in cosmological hydrodynamical simulations: How stochastic processes, numerical effects, and reproducibility limits impact predictability},
    journal = {Astron. Comput.},
    pages = {101127},
    year = {2026},
    issn = {2213-1337},
    doi = {https://doi.org/10.1016/j.ascom.2026.101127},
    url = {https://www.sciencedirect.com/science/article/pii/S2213133726000697},
    author = {Chaitra and Antonio Ragagnin and Milena Valentini and Giuseppe Murante and Stefano Borgani and Giuliano Taffoni}
}

@ARTICLE{Bonafede_2011,
       author = {{Bonafede}, A. and {Dolag}, K. and {Stasyszyn}, F. and {Murante}, G. and {Borgani}, S.},
        title = "{A non-ideal magnetohydrodynamic GADGET: simulating massive galaxy clusters}",
      journal = {\mnras},
         year = 2011,
        month = dec,
       volume = {418},
       number = {4},
        pages = {2234-2250},
          doi = {10.1111/j.1365-2966.2011.19523.x},
archivePrefix = {arXiv},
       eprint = {1107.0968},
 primaryClass = {astro-ph.CO},
       adsurl = {https://ui.adsabs.harvard.edu/abs/2011MNRAS.418.2234B}
}

@ARTICLE{TedeschiPrades_2025,
       author = {{Tedeschi-Prades}, G. and {Birnstiel}, T. and {Dolag}, K. and {Ercolano}, B. and {Hutchison}, M.},
        title = "{Modeling dust dynamics in OpenGadget3: I. SPH implementation of the One-Fluid model}",
      journal = {\aap},
         year = 2025,
        month = dec,
       volume = {704},
          eid = {A92},
        pages = {A92},
          doi = {10.1051/0004-6361/202554069},
archivePrefix = {arXiv},
       eprint = {2511.00631},
 primaryClass = {astro-ph.IM},
       adsurl = {https://ui.adsabs.harvard.edu/abs/2025A&A...704A..92T}
}

@ARTICLE{Damiano_2024,
       author = {{Damiano}, Alice and {Valentini}, Milena and {Borgani}, Stefano and {Tornatore}, Luca and {Murante}, Giuseppe and {Ragagnin}, Antonio and {Ragone-Figueroa}, Cinthia and {Dolag}, Klaus},
        title = "{Dynamical friction and the evolution of black holes in cosmological simulations: A new implementation in OpenGadget3}",
      journal = {\aap},
         year = 2024,
        month = dec,
       volume = {692},
          eid = {A81},
        pages = {A81},
          doi = {10.1051/0004-6361/202450021},
archivePrefix = {arXiv},
       eprint = {2403.12600},
 primaryClass = {astro-ph.CO},
       adsurl = {https://ui.adsabs.harvard.edu/abs/2024A&A...692A..81D}
}

@ARTICLE{Sala_2024,
       author = {{Sala}, Luca and {Valentini}, Milena and {Biffi}, Veronica and {Dolag}, Klaus},
        title = "{Supermassive black hole spin evolution in cosmological simulations with OPENGADGET3}",
      journal = {\aap},
         year = 2024,
        month = may,
       volume = {685},
          eid = {A92},
        pages = {A92},
          doi = {10.1051/0004-6361/202348925},
archivePrefix = {arXiv},
       eprint = {2312.07657},
 primaryClass = {astro-ph.GA},
       adsurl = {https://ui.adsabs.harvard.edu/abs/2024A&A...685A..92S}
}

@ARTICLE{Groth_2023,
       author = {{Groth}, Frederick and {Steinwandel}, Ulrich P. and {Valentini}, Milena and {Dolag}, Klaus},
        title = "{The cosmological simulation code OPENGADGET3 - implementation of meshless finite mass}",
      journal = {\mnras},
         year = 2023,
        month = nov,
       volume = {526},
       number = {1},
        pages = {616-644},
          doi = {10.1093/mnras/stad2717},
archivePrefix = {arXiv},
       eprint = {2301.03612},
 primaryClass = {astro-ph.IM},
       adsurl = {https://ui.adsabs.harvard.edu/abs/2023MNRAS.526..616G}
}

@ARTICLE{Fischer_2026,
       author = {{Fischer}, Moritz S. and {Wiertel}, Marc and {Arido}, Cenanda and {Patil}, Yashraj and {Ragagnin}, Antonio and {Dolag}, Klaus and {Br{\"u}ggen}, Marcus and {Garny}, Mathias and {Robertson}, Andrew and {Schmidt-Hoberg}, Kai},
        title = "{The Cosmological Simulation Code OpenGadget3 - Implementation of Self-Interacting Dark Matter}",
      journal = {arXiv e-prints},
         year = 2026,
        month = mar,
          eid = {arXiv:2603.10107},
        pages = {arXiv:2603.10107},
          doi = {10.48550/arXiv.2603.10107},
archivePrefix = {arXiv},
       eprint = {2603.10107},
 primaryClass = {astro-ph.IM},
       adsurl = {https://ui.adsabs.harvard.edu/abs/2026arXiv260310107F}
}

@ARTICLE{Fisher_2022b,
       author = {{Fischer}, Moritz S. and {Br{\"u}ggen}, Marcus and {Schmidt-Hoberg}, Kai and {Dolag}, Klaus and {Kahlhoefer}, Felix and {Ragagnin}, Antonio and {Robertson}, Andrew},
        title = "{Cosmological simulations with rare and frequent dark matter self-interactions}",
      journal = {\mnras},
         year = 2022,
        month = oct,
       volume = {516},
       number = {2},
        pages = {1923-1940},
          doi = {10.1093/mnras/stac2207},
archivePrefix = {arXiv},
       eprint = {2205.02243},
 primaryClass = {astro-ph.CO},
       adsurl = {https://ui.adsabs.harvard.edu/abs/2022MNRAS.516.1923F}
}

@ARTICLE{Fischer_2022,
       author = {{Fischer}, Moritz S. and {Br{\"u}ggen}, Marcus and {Schmidt-Hoberg}, Kai and {Dolag}, Klaus and {Ragagnin}, Antonio and {Robertson}, Andrew},
        title = "{Unequal-mass mergers of dark matter haloes with rare and frequent self-interactions}",
      journal = {\mnras},
         year = 2022,
        month = mar,
       volume = {510},
       number = {3},
        pages = {4080-4099},
          doi = {10.1093/mnras/stab3544},
archivePrefix = {arXiv},
       eprint = {2109.10035},
 primaryClass = {astro-ph.CO},
       adsurl = {https://ui.adsabs.harvard.edu/abs/2022MNRAS.510.4080F}
}

@ARTICLE{Fischer_2021,
       author = {{Fischer}, Moritz S. and {Br{\"u}ggen}, Marcus and {Schmidt-Hoberg}, Kai and {Dolag}, Klaus and {Kahlhoefer}, Felix and {Ragagnin}, Antonio and {Robertson}, Andrew},
        title = "{N-body simulations of dark matter with frequent self-interactions}",
      journal = {\mnras},
         year = 2021,
        month = jul,
       volume = {505},
       number = {1},
        pages = {851-868},
          doi = {10.1093/mnras/stab1198},
archivePrefix = {arXiv},
       eprint = {2012.10277},
 primaryClass = {astro-ph.CO},
       adsurl = {https://ui.adsabs.harvard.edu/abs/2021MNRAS.505..851F}
}

@article{Pedratti_2026,
	author = {{Pedratti, S.} and {Pizzuti, L.} and {Fossati, M.} and {Biviano, A.} and {Boselli, A.} and {Ragagnin, A.} and {Carlin, A.}},
	title = {Novel insights into the Coma cluster kinematics with DESI - I. Linking mass profile, orbital anisotropy, and galaxy populations},
	DOI= "10.1051/0004-6361/202659068",
	url= "https://doi.org/10.1051/0004-6361/202659068",
	journal = {\aap},
	year = 2026,
	volume = 710,
	pages = "A218",
}

@ARTICLE{Mamon_2013,
       author = {{Mamon}, Gary A. and {Biviano}, Andrea and {Bou{\'e}}, Gwena{\"e}l},
        title = "{MAMPOSSt: Modelling Anisotropy and Mass Profiles of Observed Spherical Systems - I. Gaussian 3D velocities}",
      journal = {\mnras},
         year = 2013,
        month = mar,
       volume = {429},
       number = {4},
        pages = {3079-3098},
          doi = {10.1093/mnras/sts565},
archivePrefix = {arXiv},
       eprint = {1212.1455},
 primaryClass = {astro-ph.CO},
       adsurl = {https://ui.adsabs.harvard.edu/abs/2013MNRAS.429.3079M}
}

@ARTICLE{Zwicky_2009,
       author = {{Zwicky}, F.},
        title = "{Republication of: The redshift of extragalactic nebulae}",
      journal = {Gen. Relativ. Gravit.},
         year = 2009,
        month = jan,
       volume = {41},
       number = {1},
        pages = {207-224},
          doi = {10.1007/s10714-008-0707-4},
       adsurl = {https://ui.adsabs.harvard.edu/abs/2009GReGr..41..207Z}
    }

@article{Duffy_2010,
    author = {Duffy, Alan R. and Schaye, Joop and Kay, Scott T. and Vecchia, Claudio Dalla and Battye, Richard A. and Booth, C. M.},
    title = {Impact of baryon physics on dark matter structures: a detailed simulation study of halo density profiles},
    journal = {\mnras},
    volume = {405},
    number = {4},
    pages = {2161-2178},
    year = {2010},
    month = {07},
    issn = {0035-8711},
    doi = {10.1111/j.1365-2966.2010.16613.x},
    url = {https://doi.org/10.1111/j.1365-2966.2010.16613.x},
    eprint = {https://academic.oup.com/mnras/article-pdf/405/4/2161/18430468/mnras0405-2161.pdf},
    }

@article{Martin_2023,
    author = {Sommer, Martin W and Schrabback, Tim and Ragagnin, Antonio and Rockenfeller, Robert},
    title = {Weak lensing mass bias and the alignment of centre proxies},
    journal = {\mnras},
    volume = {532},
    number = {3},
    pages = {3359-3374},
    year = {2024},
    month = {08},
    issn = {0035-8711},
    doi = {10.1093/mnras/stae1580},
    url = {https://doi.org/10.1093/mnras/stae1580},
    eprint = {https://academic.oup.com/mnras/article-pdf/532/3/3359/58633780/stae1580.pdf},
}

@ARTICLE{Giocoli_2024,
       author = {{Euclid Collaboration: Giocoli}, C. and {Meneghetti}, M. and {Rasia}, E. and others},
        title = "{Euclid preparation. XXXII. Evaluating the weak-lensing cluster mass biases using the Three Hundred Project hydrodynamical simulations}",
      journal = {\aap},
         year = 2024,
        month = jan,
       volume = {681},
          eid = {A67},
        pages = {A67},
          doi = {10.1051/0004-6361/202346058},
archivePrefix = {arXiv},
       eprint = {2302.00687},
 primaryClass = {astro-ph.CO},
       adsurl = {https://ui.adsabs.harvard.edu/abs/2024A&A...681A..67E}
}

@article{Meneghetti_2014,
   title={THE MUSIC OF CLASH: PREDICTIONS ON THE CONCENTRATION-MASS RELATION},
   volume={797},
   ISSN={1538-4357},
   url={http://dx.doi.org/10.1088/0004-637X/797/1/34},
   DOI={10.1088/0004-637x/797/1/34},
   number={1},
   journal={\apj},
   publisher={American Astronomical Society},
   author={Meneghetti, M. and Rasia, E. and Vega, J. and Merten, J. and Postman, M. and Yepes, G. and Sembolini, F. and Donahue, M. and Ettori, S. and Umetsu, K. and Balestra, I. and Bartelmann, M. and Benítez, N. and Biviano, A. and Bouwens, R. and Bradley, L. and Broadhurst, T. and Coe, D. and Czakon, N. and De Petris, M. and Ford, H. and Giocoli, C. and Gottlöber, S. and Grillo, C. and Infante, L. and Jouvel, S. and Kelson, D. and Koekemoer, A. and Lahav, O. and Lemze, D. and Medezinski, E. and Melchior, P. and Mercurio, A. and Molino, A. and Moscardini, L. and Monna, A. and Moustakas, J. and Moustakas, L. A. and Nonino, M. and Rhodes, J. and Rosati, P. and Sayers, J. and Seitz, S. and Zheng, W. and Zitrin, A.},
   year={2014},
   month=nov, 
   pages={34} 
   }

@ARTICLE{Becker_2011,
       author = {{Becker}, Matthew R. and {Kravtsov}, Andrey V.},
        title = "{On the Accuracy of Weak-lensing Cluster Mass Reconstructions}",
      journal = {\apj},
         year = 2011,
        month = oct,
       volume = {740},
       number = {1},
          eid = {25},
        pages = {25},
          doi = {10.1088/0004-637X/740/1/25},
archivePrefix = {arXiv},
       eprint = {1011.1681},
 primaryClass = {astro-ph.CO},
       adsurl = {https://ui.adsabs.harvard.edu/abs/2011ApJ...740...25B}
    }

@ARTICLE{Wechsler_2002,
       author = {{Wechsler}, Risa H. and {Bullock}, James S. and {Primack}, Joel R. and {Kravtsov}, Andrey V. and {Dekel}, Avishai},
        title = "{Concentrations of Dark Halos from Their Assembly Histories}",
      journal = {\apj},
         year = 2002,
        month = mar,
       volume = {568},
       number = {1},
        pages = {52-70},
          doi = {10.1086/338765},
archivePrefix = {arXiv},
       eprint = {astro-ph/0108151},
 primaryClass = {astro-ph},
       adsurl = {https://ui.adsabs.harvard.edu/abs/2002ApJ...568...52W}
    }

@article{Despali_2026,
   title={The AIDA-TNG project: Dark matter profiles and concentrations in alternative dark matter models},
   volume={708},
   ISSN={1432-0746},
   url={http://dx.doi.org/10.1051/0004-6361/202558686},
   DOI={10.1051/0004-6361/202558686},
   journal={\aap},
   publisher={EDP Sciences},
   author={Despali, Giulia and Giocoli, Carlo and Moscardini, Lauro and Pillepich, Annalisa and Vogelsberger, Mark and Meneghetti, Massimo},
   year={2026},
   month=Mar, pages={A47} 
   }

@article{Romanello_2026,
	author = {{Romanello, M.} and {Despali, G.} and {Marulli, F.} and {Giocoli, C.} and {Moscardini, L.} and {Vogelsberger, M.}},
	title = {The AIDA-TNG project: Abundance, radial distribution, and clustering properties of halos in alternative dark matter models},
	DOI= "10.1051/0004-6361/202558242",
	url= "https://doi.org/10.1051/0004-6361/202558242",
	journal = {\aap},
	year = 2026,
	volume = 707,
	pages = "A172",
}

@article{Fischer_Yu_2026,
	author = {{Fischer, Moritz S.} and {Yu, Hai-Bo}},
	title = {The dark fate of ultra-faint dwarfs: Gravothermal collapse in action},
	DOI= "10.1051/0004-6361/202659748",
	url= "https://doi.org/10.1051/0004-6361/202659748",
	journal = {\aap},
	year = 2026,
	volume = 711,
	pages = "A68",
}

@article{Wang_2026,
    doi = {10.3847/2041-8213/ae375a},
    url = {https://doi.org/10.3847/2041-8213/ae375a},
    year = {2026},
    month = {jan},
    publisher = {The American Astronomical Society},
    volume = {997},
    number = {2},
    pages = {L43},
    author = {Wang, Yi-Ying and Yang, Daneng and Lu, Keyu and Tsai, Yue-Lin Sming and Fan, Yi-Zhong},
    title = {Dark-matter-deficient Galaxies from Collisions: A New Probe of Bursty Feedback and Dark Matter Physics},
    journal = {\apjl}
}

@ARTICLE{Roberts_2025,
       author = {{Roberts}, M. Grant and {Kaplinghat}, Manoj and {Valli}, Mauro and {Yu}, Hai-Bo},
        title = "{Gravothermal collapse and the diversity of galactic rotation curves}",
      journal = {\prd},
         year = 2025,
        month = may,
       volume = {111},
       number = {10},
          eid = {103041},
        pages = {103041},
          doi = {10.1103/PhysRevD.111.103041},
archivePrefix = {arXiv},
       eprint = {2407.15005},
 primaryClass = {astro-ph.GA},
       adsurl = {https://ui.adsabs.harvard.edu/abs/2025PhRvD.111j3041R}
}

\begin{appendix}
\nolinenumbers
\onecolumn
\section{\lcdm mass fits and $M_{200}$ distributions} \label{app:mass_fits}

\begin{figure*}[ht!]
    \centering
    \includegraphics[width=\hsize]{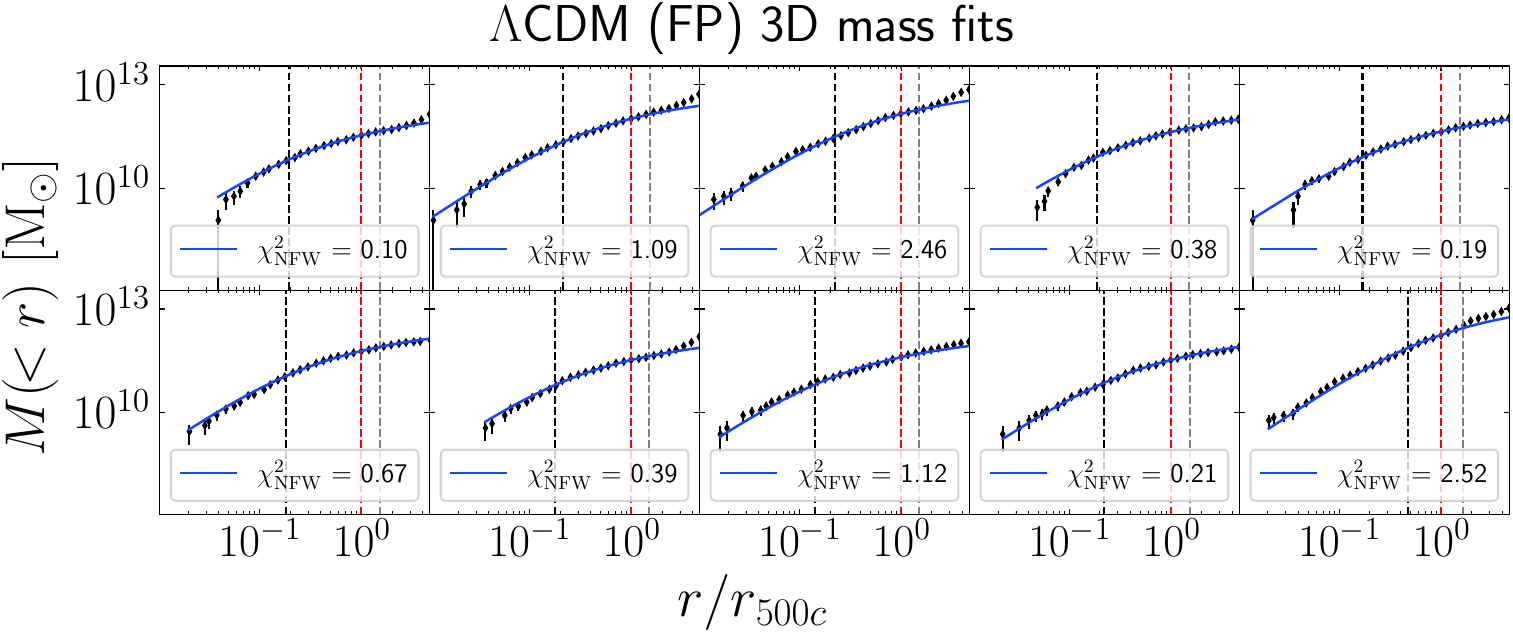}
    \caption{Ten example 3D mass profiles from the FP \lcdm sample, along with the corresponding NFW fits (solid blue line). We also show the NFW $r_{200c}$ and $r_{\rm s}$ as the dashed gray and black lines, respectively, and the halo $r_{500c}$ as the dashed red line.}
    \label{fig:mass_fit_profiles}
\end{figure*}

\begin{figure*}[ht!]
    \centering
    \includegraphics[width=0.8\hsize]{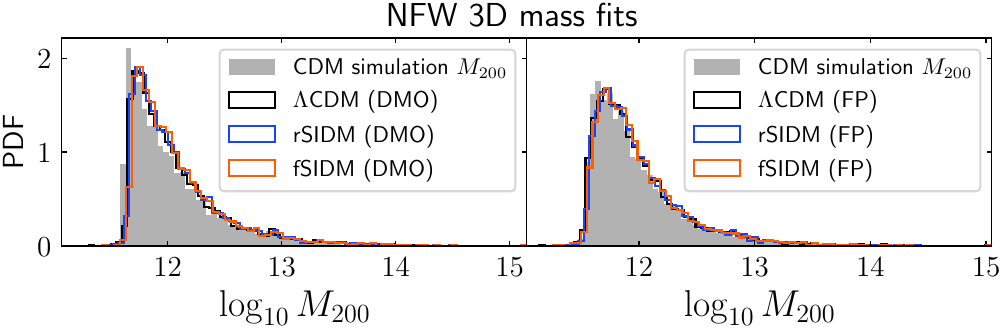}
    \caption{Halo virial mass distributions for all DMO (\textit{left}) and FP (\textit{right}) simulations. We compare the CDM mass distribution obtained from particle estimates (gray histogram) with the one obtained from fitting our halos with an NFW model. Overall, both methods resulted in similar distributions, without any significant deviations.}
    \label{fig:M200_distr}
\end{figure*}

In Fig.~\ref{fig:mass_fit_profiles} we show ten randomly selected FP CDM three-dimensional mass profiles along with their best-fit NFW model. We also show, in Fig.~\ref{fig:M200_distr}, the distribution of best-fit 3D NFW $M_{200}$ compared to those obtained from halo finder estimates for each DM model: overall both distributions are similar, without particularly significant deviations in either the DMO or FP case.

\twocolumn

\onecolumn
\section{Estimation of the gravothermal evolution for our SIDM halos with $M_{200} \gtrsim 5 \times 10^{12}\ {\rm M_\odot}$} \label{app:grav_evolution}

In this section, we estimate the central halo density $\rho_{\rm c}$ of our FP SIDM halos as a function of their age $t_{\rm age}$, in order to provide an estimate for the state of their gravothermal evolution. In particular, we compared our central densities with the theoretical model by~\citep{Yang_2023}, which describes the evolution of an isolated NFW SIDM halo central density using a single parameter $\beta$, provided a prior appropriate rescaling of the densities and timescales. Therefore, we characterized the evolution of our halos through a rescaled central density $\hat \rho_{\rm c} \equiv \rho_{\rm c} / \rho_{\rm 0}$ as a function of rescaled time $\beta \hat \sigma \hat t$, where $\rho_0$ is the NFW characteristic density, $\hat t \equiv t_{\rm age} / \sqrt{4 \uppi \mathrm{G} \rho_0}$ and $\hat \sigma$ is the effective SIDM cross section acting on the halo core (see, e.g.,~\citealt{Yang_2022, Yang_2023} for more details).

Likewise ~\cite{Yang_2023}, we fixed $\beta = 1$ to provide a rough indication of the halo gravothermal evolution. We assumed the same $t_{\rm age} = 9\ {\rm Gyr}$ for all of our halos, thus providing an upper limit to the central density of each halo.
We computed the cross section as weighted with the fifth power of the velocity and assuming that the velocities are well described by a Maxwell–Boltzmann distribution with a characteristic velocity $V_{\rm max}/\sqrt{3}$~\citep[see Eq. 4.2 in][]{Yang_2022}, so that  $\sigma = 3 \langle v^5 \sigma_v(v) \rangle / \langle v^5 \rangle / 2$, where $\sigma_v$ is the viscosity cross section as defined in Sect.~\ref{sec:SIDM_models}.

\begin{figure}[ht!]
    \centering
    \includegraphics[width=0.65\hsize]{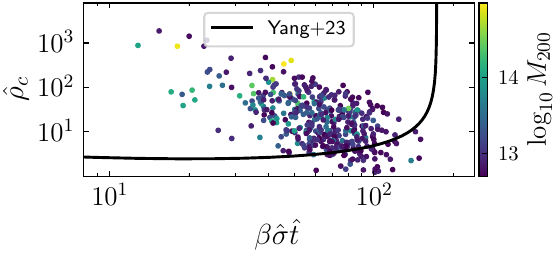}
    \caption{Upper limit on the rescaled central density over rescaled time of our FP SIDM halos, computed assuming an age of $9$ Gyr for all halos. We only considered halos with $M_{200} \geq 5 \times 10^{12}\ {\rm M_\odot}$, and all points are colored based on their virial mass. The black line represents the theoretical evolution model from~\cite{Yang_2023}.}
    \label{fig:SIDM_evolution}
\end{figure}

The resulting rescaled densities are shown in Fig.~\ref{fig:SIDM_evolution}, along with the theoretical model of~\citep{Yang_2023}: while most halos exhibit an enhanced density compared to the model due to the presence of baryons, it is clear that all halos lie before the core-collapse phase, with less massive halos presenting a more advanced evolution stage than more massive ones, as expected from SIDM interactions affecting more prominently the smaller, more concentrated halos (see Fig.~\ref{fig:SIDM_cross}).
In addition, we note that the evolution of a DM halo within the cosmological context is more complicated: as shown in \citep{Silverman_2026}, halo evolution can be strongly affected by mergers, preventing or slowing the gravothermal collapse.

\twocolumn

\onecolumn
\section{3D and 2D stacked density profiles} \label{app:density_profiles}
\begin{figure*}[ht!]
    \centering
    \centerline{
    \includegraphics[width=0.5\hsize]{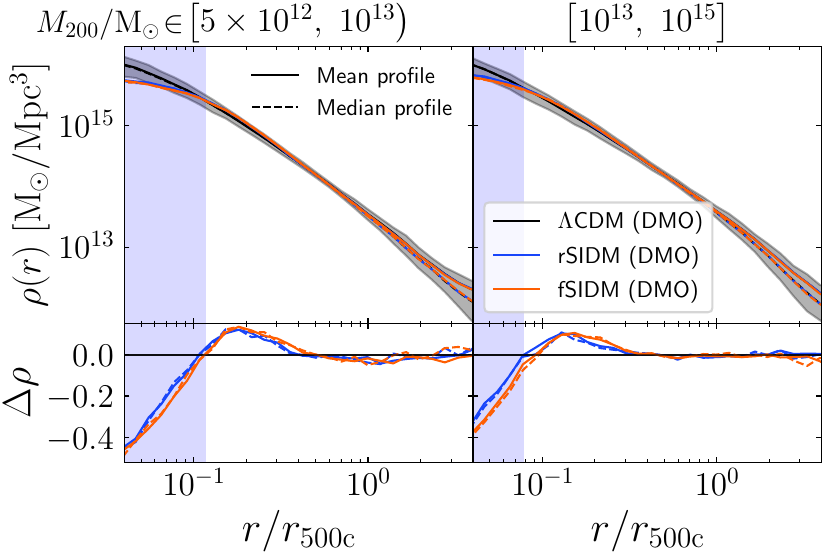}
    \includegraphics[width=0.5\hsize]{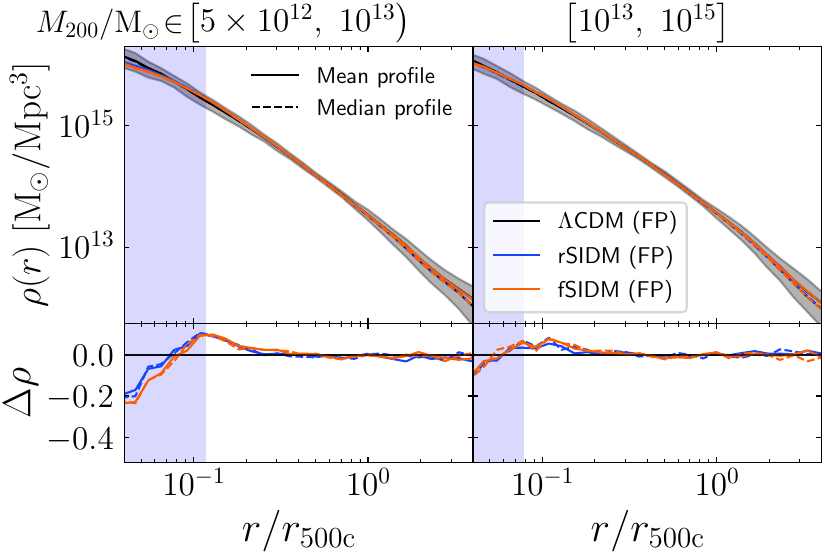}}
    \centerline{
    \includegraphics[width=0.5\hsize]{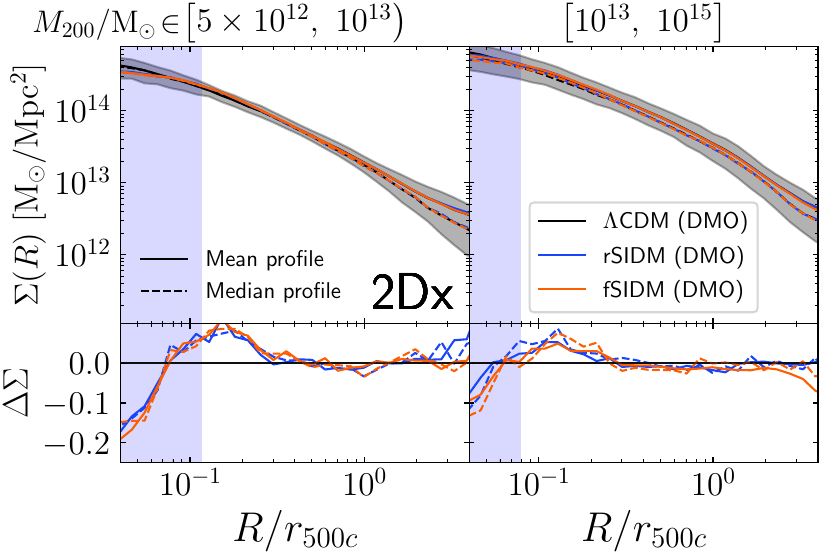}
    \includegraphics[width=0.5\hsize]{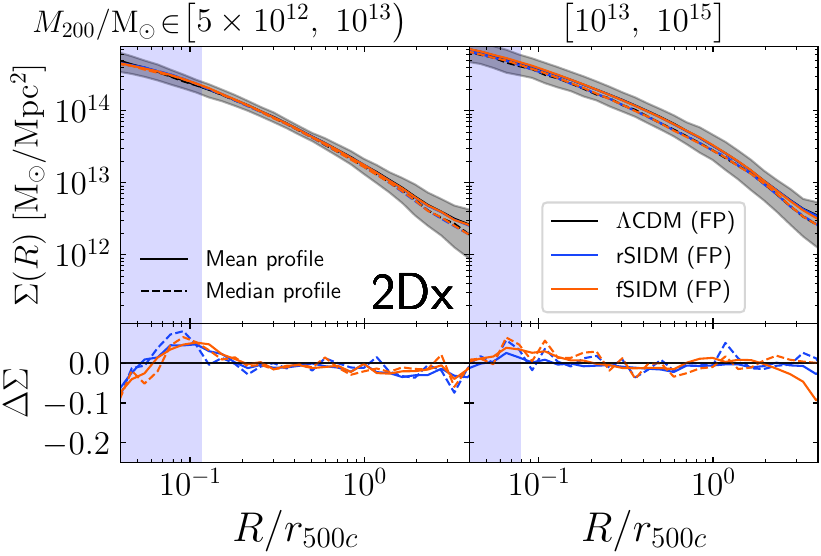}}
    \caption{Stacked 3D (\textit{top}) and 2Dx (\textit{bottom}) DM density profiles for the DMO (\textit{left}) and FP (\textit{right}) simulations. The color code is the same as in Fig.~\ref{fig:mass_profiles}. The bottom panel shows the relative differences between each model for both mean and median profiles. We also define, for a given quantity $X$, $\Delta X \equiv (X - X_{\rm CDM}) / X_{\rm CDM}$.}
    \label{fig:den_profiles}
\end{figure*}

Figure~\ref{fig:den_profiles} shows the stacked halo density profiles and their relative differences, both 3D and projected, in the DMO and FP cases. We only show the projected profiles along the $x$-direction as all other projections yield similar results. Overall the SIDM density profiles exhibit a consistent behavior with their associated mass profiles in Fig.~\ref{fig:mass_profiles}.

\twocolumn

\onecolumn
\section{Gas and stellar mass profiles of the D10, D15, and D16 region clusters compared to MACS~J1206 and Abell~S1063} \label{app:gas_star_profiles}
\begin{figure*}[ht!]
    \centering
    \centerline{
    \includegraphics[width=0.5\hsize]{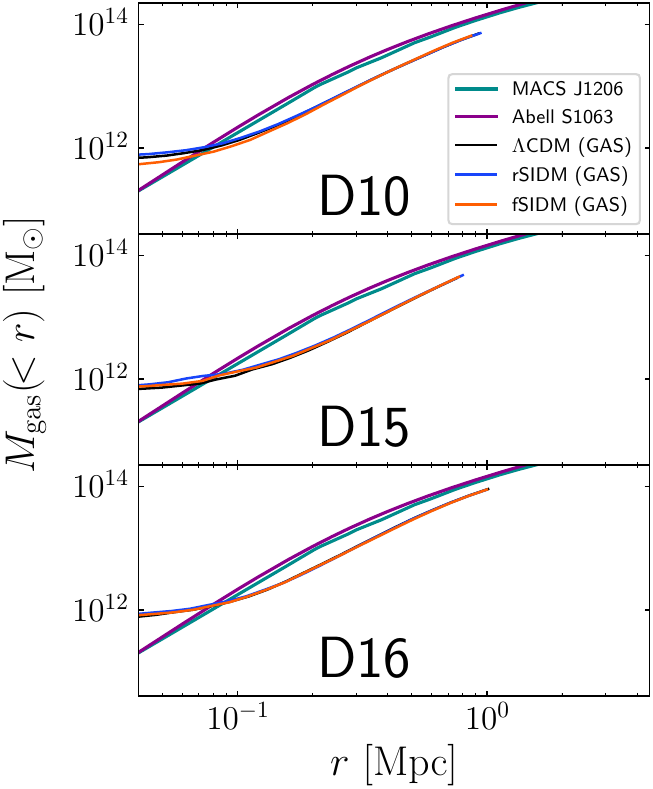}
    \includegraphics[width=0.5\hsize]{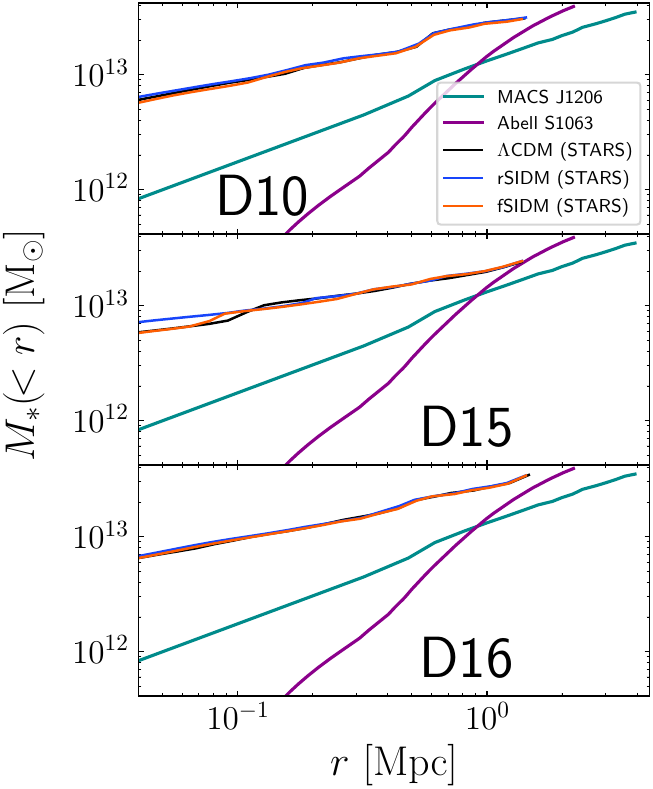}}
    \caption{Comparison of the gas (\textit{left}) and stellar (\textit{right}) mass profiles of our simulated cluster-sized halos belonging to the D10 (\textit{top}), D15 (\textit{middle}) and D16 (\textit{bottom}) regions, and the real galaxy clusters MACS~J1206 (the blue line;~\citealt{Biviano_2023}) and Abell~S1063 (the purple line;~\citealt{Sartoris_2020}). The observed profile uncertainties are small and are not visible in the plots.}
    \label{fig:gas_star_profs}
\end{figure*}

In Fig.~\ref{fig:gas_star_profs} we display the mass profiles of the gas and stellar components corresponding to the most massive simulated clusters in the D10, D15 and D16 DIANOGA-SIDM regions -- for all of our DM models -- compared to those belonging to the MACS~J1206 and Abell~S1063 clusters inferred from galaxy kinematic analyses (see Sect.~\ref{sec:obs_cluster}).

\twocolumn

\onecolumn
\section{NFW concentration-mass relation in all cosmologies} \label{app:cM_relations}
\begin{figure*}[ht!]
    \centering
    \includegraphics[width=\hsize]{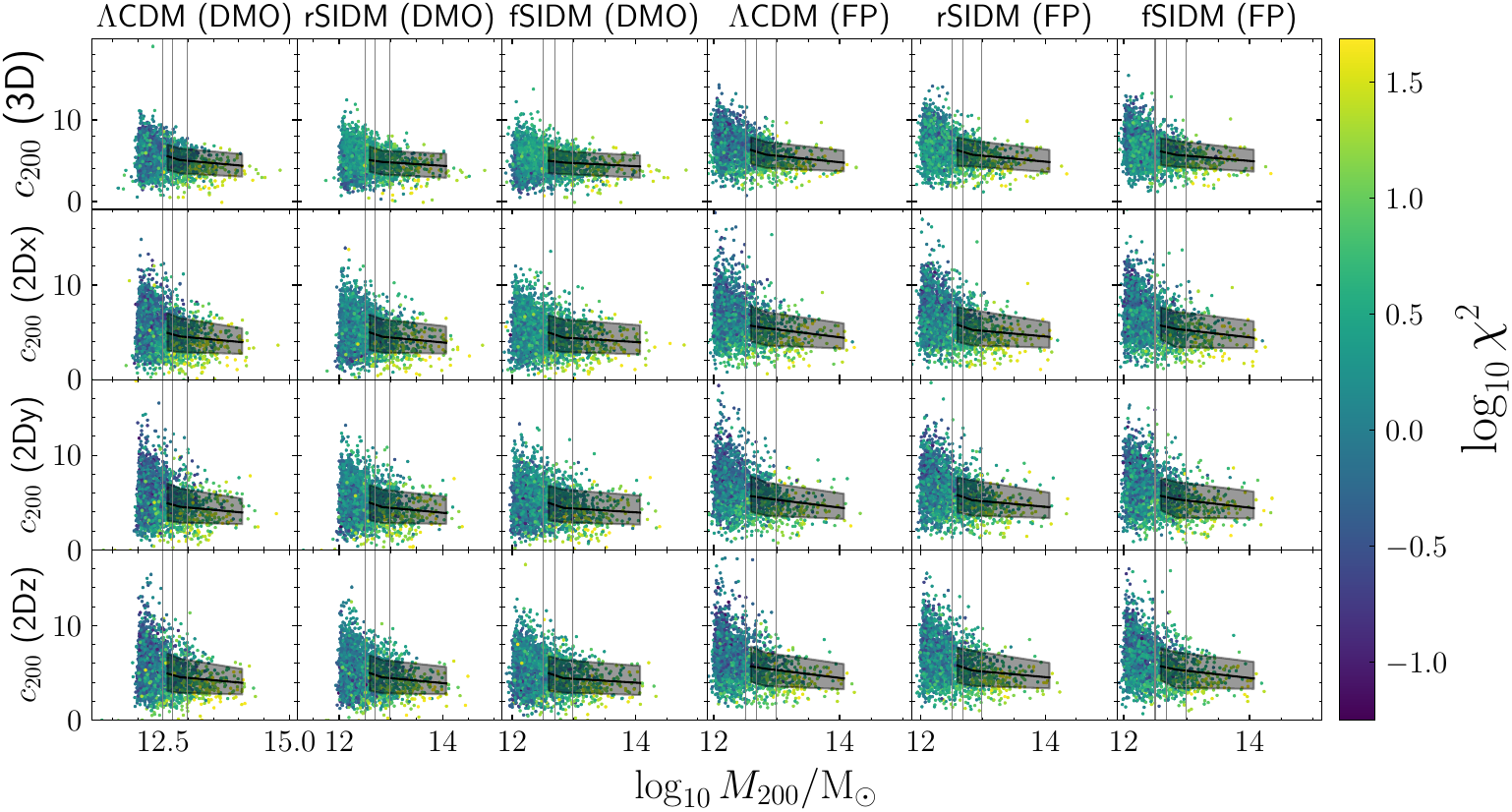}
    \caption{Halo c-M relations for all DMO and FP models. The concentrations and virial masses have been obtained from the NFW fits as described in Sect.~\ref{sec:c_m_relation}, for both 3D and projected cases. All points are colored based on their NFW fit chi-square per degree of freedom. Shown is also the median concentration and its $68\%$ scatter (the black line) computed across three logarithmic mass bins (whose edges are indicated by the vertical gray lines).}
    \label{fig:cM_relation}
\end{figure*}

In Fig.~\ref{fig:cM_relation} we display the NFW c-M relation computed for every DM model, and for both 3D and projected cases. We further computed the median concentration across three equally populated mass bins, and only considering halos with $M_{200} \geq 5 \times 10^{12}\ \mathrm{M_\odot}$ due to simulation resolution concerns.

\twocolumn

\end{appendix}

\end{document}